\newcount\mgnf  
\mgnf=0

\ifnum\mgnf=0
\magnification=1000   
\hsize=15truecm\vsize=22.5truecm
   \parindent=0.3cm\baselineskip=0.45cm\fi
\ifnum\mgnf=1
   \magnification=\magstephalf
   \baselineskip=18truept plus0.1pt minus0.1pt \parindent=0.9truecm
\fi

\ifnum\mgnf=2\magnification=1200\fi

\ifnum\mgnf=0
\def\openone{\leavevmode\hbox{\ninerm 1\kern-3.3pt\tenrm1}}%
\def\*{\vskip3.truemm}\fi
\ifnum\mgnf=1
\def\openone{\leavevmode\hbox{\ninerm 1\kern-3.63pt\tenrm1}}%
\def\*{\vglue0.3truecm}\fi
\ifnum\mgnf=2\def\*{\vglue0.7truecm}\fi
\openout15=\jobname.aux

\font\titolo=cmbx12\font\titolone=cmbx10 scaled\magstep 2%
\font\sc=cmcsc10\font\css=cmcsc8%
\font\indbf=cmbx10 scaled\magstep2
\font\ottorm=cmr8\font\ninerm=cmr9%
\font\msytw=msbm9 scaled\magstep1%
\font\msytwww=msbm5 scaled\magstep1%
%
%
%
\def\st{\scriptstyle}%
%


\font\tenmib=cmmib10 \font\eightmib=cmmib8
\font\sevenmib=cmmib7\font\fivemib=cmmib5 
\font\ottoit=cmti8
\font\fiveit=cmti5\font\sixit=cmti6
\font\fivei=cmmi5\font\sixi=cmmi6\font\ottoi=cmmi8
\font\ottorm=cmr8
\font\ottosy=cmsy8\font\sixsy=cmsy6\font\fivesy=cmsy5
\font\ottobf=cmbx8\font\sixbf=cmbx6\font\fivebf=cmbx5%
\font\ottocss=cmcsc8%

\def\ottopunti{\def\rm{\fam0\ottorm}\def\it{\fam6\ottoit}%
\def\bf{\fam7\ottobf}%
\textfont1=\ottoi\scriptfont1=\sixi\scriptscriptfont1=\fivei%
\textfont2=\ottosy\scriptfont2=\sixsy\scriptscriptfont2=\fivesy%
\textfont4=\ottocss\scriptfont4=\sc\scriptscriptfont4=\sc%
\textfont5=\eightmib\scriptfont5=\sevenmib\scriptscriptfont5=\fivemib%
\textfont6=\ottoit\scriptfont6=\sixit\scriptscriptfont6=\fiveit%
\textfont7=\ottobf\scriptfont7=\sixbf\scriptscriptfont7=\fivebf%
\setbox\strutbox=\hbox{\vrule height7pt depth2pt width0pt}%
\normalbaselineskip=9pt\rm}

\textfont5=\tenmib\scriptfont5=\sevenmib\scriptscriptfont5=\fivemib
\mathchardef\Ba   = "050B  
\mathchardef\Bb   = "050C  
\mathchardef\Bg   = "050D  
\mathchardef\Bd   = "050E  
\mathchardef\Be   = "0522  
\mathchardef\Bee  = "050F  
\mathchardef\Bz   = "0510  
\mathchardef\Bh   = "0511  
\mathchardef\Bthh = "0512  
\mathchardef\Bth  = "0523  
\mathchardef\Bi   = "0513  
\mathchardef\Bk   = "0514  
\mathchardef\Bl   = "0515  
\mathchardef\Bm   = "0516  
\mathchardef\Bn   = "0517  
\mathchardef\Bx   = "0518  
\mathchardef\Bom  = "0530  
\mathchardef\Bp   = "0519  
\mathchardef\Br   = "0525  
\mathchardef\Bro  = "051A  
\mathchardef\Bs   = "051B  
\mathchardef\Bsi  = "0526  
\mathchardef\Bt   = "051C  
\mathchardef\Bu   = "051D  
\mathchardef\Bf   = "0527  
\mathchardef\Bff  = "051E  
\mathchardef\Bch  = "051F  
\mathchardef\Bps  = "0520  
\mathchardef\Bo   = "0521  
\mathchardef\Bome = "0524  
\mathchardef\BG   = "0500  
\mathchardef\BD   = "0501  
\mathchardef\BTh  = "0502  
\mathchardef\BL   = "0503  
\mathchardef\BX   = "0504  
\mathchardef\BP   = "0505  
\mathchardef\BS   = "0506  
\mathchardef\BU   = "0507  
\mathchardef\BF   = "0508  
\mathchardef\BPs  = "0509  
\mathchardef\BO   = "050A  
\mathchardef\BDpr = "0540  
\mathchardef\Bstl = "053F  

%
%
%
%
%
%
%

\global\newcount\numsec\global\newcount\numapp
\global\newcount\numfor\global\newcount\numfig
\global\newcount\numsub
\numsec=0\numapp=0\numfig=0
\def\veroparagrafo{\number\numsec}\def\veraformula{\number\numfor}
\def\veraappendice{\number\numapp}\def\verasub{\number\numsub}
\def\verafigura{\number\numfig}

\def\Section(#1,#2){\advance\numsec by 1\numfor=1\numsub=1\numfig=1%
\SIA p,#1,{\veroparagrafo} %
\write15{\string\Fp (#1){\secc(#1)}}%
\write16{ sec. #1 ==> \secc(#1)  }%
\0\hbox
{\titolo\hfill
\number\numsec. #2\hfill%
\expandafter{\hglue-1truecm\alato(sec. #1)}}}

\def\appendix(#1,#2){\advance\numapp by 1\numfor=1\numsub=1\numfig=1%
\SIA p,#1,{A\veraappendice} %
\write15{\string\Fp (#1){\secc(#1)}}%
\write16{ app. #1 ==> \secc(#1)  }%
\hbox to \hsize{\titolo Appendix A\number\numapp. #2\hfill%
\expandafter{\alato(app. #1)}}\*%
}

\def\senondefinito#1{\expandafter\ifx\csname#1\endcsname\relax}

\def\SIA #1,#2,#3 {\senondefinito{#1#2}%
\expandafter\xdef\csname #1#2\endcsname{#3}\else
\write16{???? ma #1#2 e' gia' stato definito !!!!} \fi}

\def \Fe(#1)#2{\SIA fe,#1,#2 }
\def \Fp(#1)#2{\SIA fp,#1,#2 }
\def \Fg(#1)#2{\SIA fg,#1,#2 }

\def\etichetta(#1){(\veroparagrafo.\veraformula)%
\SIA e,#1,(\veroparagrafo.\veraformula) %
\global\advance\numfor by 1%
\write15{\string\Fe (#1){\equ(#1)}}%
\write16{ EQ #1 ==> \equ(#1)  }}

\def\etichettaa(#1){(A\veraappendice.\veraformula)%
\SIA e,#1,(A\veraappendice.\veraformula) %
\global\advance\numfor by 1%
\write15{\string\Fe (#1){\equ(#1)}}%
\write16{ EQ #1 ==> \equ(#1) }}

\def\getichetta(#1){\veroparagrafo.\verafigura%
\SIA g,#1,{\veroparagrafo.\verafigura} %
\global\advance\numfig by 1%
\write15{\string\Fg (#1){\graf(#1)}}%
\write16{ Fig. #1 ==> \graf(#1) }}

\def\etichettap(#1){\veroparagrafo.\verasub%
\SIA p,#1,{\veroparagrafo.\verasub} %
\global\advance\numsub by 1%
\write15{\string\Fp (#1){\secc(#1)}}%
\write16{ par #1 ==> \secc(#1)  }}

\def\Eq(#1){\eqno{\etichetta(#1)\alato(#1)}}
\def\eq(#1){\etichetta(#1)\alato(#1)}
\def\Eqa(#1){\eqno{\etichettaa(#1)\alato(#1)}}
\def\eqa(#1){\etichettaa(#1)\alato(#1)}
\def\eqg(#1){\getichetta(#1)\alato(fig. #1)}
\def\sub(#1){\0\palato(p. #1){\bf \etichettap(#1).}}
\def\asub(#1){\0\palato(p. #1){\bf \etichettapa(#1).}}
\def\apprif(#1){\senondefinito{e#1}%
\eqv(#1)\else\csname e#1\endcsname\fi}

\def\equv(#1){\senondefinito{fe#1}$\clubsuit$#1%
\write16{eq. #1 non e' (ancora) definita}%
\else\csname fe#1\endcsname\fi}
\def\grafv(#1){\senondefinito{fg#1}$\clubsuit$#1%
\write16{fig. #1 non e' (ancora) definito}%
\else\csname fg#1\endcsname\fi}
\def\secv(#1){\senondefinito{fp#1}$\clubsuit$#1%
\write16{par. #1 non e' (ancora) definito}%
\else\csname fp#1\endcsname\fi}

\def\eqo{{\global\advance\numfor by 1}}
\def\equ(#1){\senondefinito{e#1}\equv(#1)\else\csname e#1\endcsname\fi}
\def\graf(#1){\senondefinito{g#1}\grafv(#1)\else\csname g#1\endcsname\fi}
\def\figura(#1){{\css Figura} \getichetta(#1)}
\def\secc(#1){\senondefinito{p#1}\secv(#1)\else\csname p#1\endcsname\fi}
\def\sec(#1){{\secc(#1)}}
\def\refe(#1){{[\secc(#1)]}}

\def\BOZZA{
\def\alato(##1){\rlap{\kern-\hsize\kern-.5truecm{$\scriptstyle##1$}}}
\def\palato(##1){\rlap{\kern-.5truecm{$\scriptstyle##1$}}}
}

\def\alato(#1){}
\def\galato(#1){}
\def\palato(#1){}


{\count255=\time\divide\count255 by 60 \xdef\hourmin{\number\count255}
        \multiply\count255 by-60\advance\count255 by\time
   \xdef\hourmin{\hourmin:\ifnum\count255<10 0\fi\the\count255}}

\def\oramin{\hourmin }

\def\data{\number\day/\ifcase\month\or gennaio \or febbraio \or marzo \or
aprile \or maggio \or giugno \or luglio \or agosto \or settembre
\or ottobre \or novembre \or dicembre \fi/\number\year;\ \oramin}
\setbox200\hbox{$\scriptscriptstyle \data $}

\newdimen\xshift \newdimen\xwidth \newdimen\yshift \newdimen\ywidth

\def\ins#1#2#3{\vbox to0pt{\kern-#2\hbox{\kern#1 #3}\vss}\nointerlineskip}

\def\eqfig#1#2#3#4#5{
\par\xwidth=#1 \xshift=\hsize \advance\xshift
by-\xwidth \divide\xshift by 2
\yshift=#2 \divide\yshift by 2%
{\hglue\xshift \vbox to #2{\vfil
#3 \includegraphics{#4.ps}
}\hfill\raise\yshift\hbox{#5}}}

\def\8{\write12}

\openin13=#1.aux \ifeof13 \relax \else
\input #1.aux \closein13\fi
\openin14=\jobname.aux \ifeof14 \relax \else
\input \jobname.aux \closein14 \fi
\immediate\openout15=\jobname.aux


\let\a=\alpha   \let\g=\gamma  \let\d=\delta \let\e=\varepsilon
\let\z=\zeta     \let\th=\theta  \let\l=\lambda
\let\m=\mu    \let\n=\nu    \let\x=\xi     \let\p=\pi    \let\r=\rho
 \let\t=\tau   \let\f=\varphi 
\let\ch=\chi     
 \let\D=\Delta  \let\L=\Lambda

\def\\{\hfill\break} \let\==\equiv

\let\io=\infty 
\def\ap{{\it a priori\ }}
\let\0=\noindent

\def\media#1{{\langle#1\rangle}}
\def\ie{\hbox{\it i.e.\ }}\def\eg{\hbox{\it e.g.\ }}
\let\dpr=\partial

\def\tende#1{\,\vtop{\ialign{##\crcr\rightarrowfill\crcr
 \noalign{\kern-1pt\nointerlineskip} \hskip3.pt${\scriptstyle
 #1}$\hskip3.pt\crcr}}\,}
\def\circage{\lower2pt\hbox{$\,\buildrel > \over {\scriptstyle \sim}\,$}}
\def\otto{\,{\kern-1.truept\leftarrow\kern-5.truept\to\kern-1.truept}\,}
\def\fra#1#2{{#1\over#2}}

\def\EE{{\cal E}}

\def\RR{{\cal R}}  
\def\DD{{\cal D}}  
 \def\QQ{{\cal Q}} 

\def\T#1{{#1_{\kern-3pt\lower7pt\hbox{$\widetilde{}$}}\kern3pt}}
\def\VVV#1{{\underline #1}_{\kern-3pt
\lower7pt\hbox{$\widetilde{}$}}\kern3pt\,}
\def\W#1{#1_{\kern-3pt\lower7.5pt\hbox{$\widetilde{}$}}\kern2pt\,}

\def\lis{\overline}
\def\etc{{\it etc}} 
  
\def\indica{\leaders \hbox to 0.5cm{\hss.\hss}\hfill}
\def\guida{\leaders\hbox to 1em{\hss.\hss}\hfill}

\def\qed{\raise1pt\hbox{\vrule height5pt width5pt depth0pt}}

\def\indic{\hbox{\raise-2pt \hbox{\indbf 1}}}

\def\RRR{\hbox{\msytw R}}

 \def\ZZZ{\hbox{\msytw Z}}
 \def\zzz{\hbox{\msytwww Z}}

\def\defi{\,{\buildrel def\over=}\,}

\def\rhs{{\it r.h.s.}\ }

\def\sqr#1#2{{\vcenter{\vbox{\hrule height.#2pt%
        \hbox{\vrule width.#2pt height#1pt \kern#1pt%
          \vrule width.#2pt}%
        \hrule height.#2pt}}}}

\def\ig{\int}

\footline={\rlap{\hbox{\copy200}}\tenrm\hss \number\pageno\hss}
\def\V#1{{\bf#1}}

\def\fiat{}

\def\asint#1{\,\vtop{\ialign{##\crcr\hss $\simeq$\hss\crcr
 \noalign{\kern-1pt\nointerlineskip} \hskip3.pt${\scriptstyle
 #1}$\hskip3.pt\crcr}}\,}

\fiat

\centerline{\titolone Constructive Quantum Field Theory}
\*

\centerline
{\it Giovanni Gallavotti}

\centerline
{\it I.N.F.N. Roma 1, Fisica Roma1}

\kern5truemm
\Section(1, Euclidean Quantum Fields)
\*

The construction of a relativistic quantum field is still an open
problem for fields in space-time dimension $d\ge 4$. The conceptual
difficulty that sometimes led to fear an incompatibility between
nontrivial quantum systems and special relativity has however been
solved in the case of dimension $d=2,3$ although, so far, has not
influenced the corresponding debate on the foundations of quantum
mechanics, still much alive.

It began in the early 1960's with Wightman's work on the axioms and
the attempts at understanding the mathematical aspects of
renormalization theory and with Hepps' renormalization theory for
scalar fields. The breakthrough idea was, perhaps, Nelson's
realization that the problem could really be studied in {\it Euclidean
form}. A solution in dimensions $d=2,3$ has been obtained in the
1960's and 1970's through a remarkable series of papers by Nelson,
Glimm, Jaffe, Guerra. While the works of Nelson and Guerra relied on
the ``Euclidean approach'' (see below) and on $d=2$ the early works of
Glimm and Jaffe dealt with $d=3$ making use of the ``Minkowskian
approach'' (based on second quantization) but making already use of a
{\it multiscale analysis} technique. The latter received great
impulsion and systematization by the adoption of Wilson's views and
methods on renormalization: in Physics terminology {\it
renormalization group} methods; a point of view taken here following
the Euclidean approach. The solution dealt initially with {\it scalar
fields} but it has been subsequently considerably extended.

The {\it Euclidean approach} studies quantum fields through the
following problems
\*

\0(1) existence of the functional integrals defining the generating
    functions of the probability distribution of the interacting
    fields in finite volume: the {\it ultraviolet stability
    problem},

\0(2) existence of the infinite volume limit of the generating
    functions: the {\it infrared problem},

\0(3) check that the infinite volume generating functions satisfy the
    axioms needed to pass from the Euclidean, probabilitstic,
    formulation to a Minkowskian formulation guaranteeing existence of
    the Hamiltonian operator, relativistic covariance, Ruelle--Haag
    scattering theory: the {\it reconstruction problem}.
\*

The characteristic problem for the construction of quantum fields is
(1) and here attention will be confined to it with the further
restriction to the paradigmatic massive scalar fields cases. The
dimension $d$ of the space-time will be $d=2,3$ unless specified
otherwise.

Given a cube $\L$ of side $L$, $\L\subset \RRR^d $,
consider the following {\it functional integral}
on the space of the fields on $\L$, \ie
on functions $\f^{(\le N)}_\Bx$ defined for $\Bx\in\L$,

$$Z_N(\L,f)=
\ig e^{-\ig_\L (\l_N
\f^{(\le N)\, 4}_\Bx +\m_N \f^{(\le N)\,2}_\Bx+\n_N+
f_\Bx \f^{(\le N)}_\Bx ) \,d\Bx}\,
P_N(d\f^{(\le N)})
\Eq(1.1)$$
The fields $\f^{(\le N)}_\Bx$ are called ``Euclidean'' fields with
{\it ultraviolet cut--off} $N>0$, $f_\Bx$ is a smooth function with
compact support bounded by $|f_\Bx|\le1$ (for definiteness), the
constants $\l_N>0$, $\m_N,\n_N$ are called {\it bare couplings}, and
$P_N$ is a Gaussian probability distribution defining the {\it free
field distribution} with {\it mass $m$} and {ultraviolet cut--off}
$N$; the probability distribution $P_N$ is determined by its
``covariance'' $C^{(\le N)}_{\Bx,\Bh}\defi\ig\f^{(\le N)}_\Bx\f^{(\le
N)}_\Bh \,d P_N$, which in the Physics literature is called a {\it
propagator}, given by

$$C^{(\le N)}_{\Bx,\Bh}=\fra1{(2\p)^d}\sum_{\V n\in\zzz^d}
\ig \fra{ e^{i\V p\cdot(\Bx-\Bh+\V n L)}}{\V p^2+m^2}\,
\ch_N(|\V p|)\,d^d \V p\Eq(1.2)$$
The sum over the integers $\V n\in\ZZZ^d$ is introduced so that the
field $\f^{(\le N)}_\Bx$ is {\it periodic} over the box $\L$: this is
not really necessary as in the limit $L\to\io$ either translation invariance
would be recovered or lack of it properly understood, but it makes the
problem more symmetric and generates a few technical simplifications;
here $\ch_N(z)$ is a {\it regularizer} and a standard choice is
$\ch_N(|\V p|)=\fra{m^2\,(\g^{2N}-1)}{\V p^2+\g^{2N} m^2}$ with $\g>1$,
which is such that

$$\fra{\ch_N(|\V p|)}{\V p^2+m^2}\=\fra1{\V p^2+m^2}-\fra1{\V
p^2+\g^{2N}m^2}\=\sum_{h=1}^N
\big(\fra1{\V p^2+\g^{2(h-1)}m^2}-\fra1{\V
p^2+\g^{2h}m^2}\big)\Eq(1.3)$$
here $\g>1$ can be chosen arbitrarily: so $\g=2$.  If $d>3$ the above
regularization will not be sufficient and a $\ch_N$ decayng faster
than $\V p^{-2}$ would be needed.

A simple estimate
yields, if $\e\in(0,1)$ is fixed and $c$ is suitably chosen,

$$|C^{(\le N)}_{\Bx,\Bh}|\le \,c\,
\g^{(d-2)N}{e^{-m|\Bx-\Bh|}}
\qquad |C^{(\le N)}_{\Bx,\Bh}-C^{(\le
N)}_{\Bx,\Bh'}|\le\, c\,\g^{(d-2)N}\,(
\g^{N} m\,|\Bh-\Bh'|)^{\e}\Eq(1.4)$$
with $\g^{(d-2) N}$ interpreted as $N$ if $d=2$.

The $\z(f)=\log\fra{Z_N(\L,f)}{Z_N(\L,0)}$ defines a ``generating
function'' of a probability distribution $P_{int}$ over the fields on
$\L$ which will be called the ``distribution with $\f^4$-interaction''
regularized on $\L$ and at length scale $m^{-1}\g^{-N}$: the integral,
in \equ(1.1),

$$V_N(\f^{(\le N)})\defi -\ig_\L (\l_N
\f^{(\le N)\, 4}_\Bx +\m_N \f^{(\le N)\,2}_\Bx+\n_N+
f_\Bx \f^{(\le N)}_\Bx ) \,d^d\Bx\Eq(1.5)$$
will be called the {\it interaction potential} with external field
$f$.  The regularization is introduced to guarantee that the integral
\equ(1.1), $\ig e^{V_N} dP_N$, is well defined if $\l_N>0$. The
momenta of $P_{int}$ are the functional derivatives of $\z(f)$: they
are called {\it Schwinger functions}.

The problem (1) can now be made precise: it is to show existence of
$\l_N,\m_N,\n_N$ so that the limit $\lim_{N\to\io}
\fra{Z_N(\L,f)}{Z_N(\L,0)}$ exists for all $f$ and {\it is not Gaussian},
\ie it is not the exponential of a quadratic form in $f$: which would
be the case if $\l_N,\m_N\to0$ fast enough: the last requirement is of
course essential because the Gaussian case describes, in the physical
interpretation, free fields and non interacting particles \ie it is
{\it trivial}. Note that $\n_N$ does not play a role: its introduction
is useful to be able to study separately the numerator and the
denominator of the fraction $\fra{Z_N(\L,f)}{Z_N(\L,0)}$.
\*
\0{\it References:} [WG65],[SW64],[Ne66],[OS73],[Si74].
\*

\Section(2, The regularized free field)
\*

Since the propagator decays exponentially over a scale $m^{-1}$ and is
smooth over a scale $m^{-1}\g^{-N}$ the fields $\f^{(\le N)}_\Bx$
sampled with distribution $P_N$ are rather singular objects. Their
properties cannot be described by a single length scale: they are
extremely large for large $N$, take independent values only beyond
distances of order $m^{-1}$ but, at the same time, they look smooth
only on the much smaller scale $m^{-1}\g^{-N}$.  Their essential
feature is that fixed $\e<1$, \eg $\e=\fra12$, with $P_N$-probability
$1$ there is $B>0$ such that (interpreting $\g^{\fra{d-2}2N}$ as $N$
if $d=2$)

$$ |\f^{(\le N)}_\Bx|\le B \g^{\fra{d-2}2N},\quad
 |\f^{(\le N)}_\Bx-\f^{(\le N)}_\Bh|< B
 \g^{\fra{d-2}2N}(\g^Nm|\Bx-\Bh|)^{\fra\e2}
\Eq(2.1)$$
and furthermore the probability of the relations in \equ(2.1) will be
$N$-independent, \ie $\f^{(\le N)}_\Bx$ are bounded and roughly of
size $\g^{\fra{d-2}2N}$ as $N\to\io$ and, on a very small length scale
$m^{-1}\g^{-N}$, almost constant.

Substantial control on the field $\f^{(\le N)}_\Bx$ statistically
sampled with distribution $P_N$ can be obtained by decomposing it,
through \equ(1.3), into ``components of various scales'': \ie as a sum
of statistically mutually {\it independent} fields whose properties
are entirely characterized by a {\it single scale of length}. This
means that they have size of order $1$ and are {\it independent and
smooth on the same length scale}.

Assuming the side of $\L$ to be an integer multiple of $m^{-1}$, let
$\QQ_h$ be a {\it pavement} of $\L$ into boxes of side $m^{-1}\g^{-h}$,
imagined {\it hierarchically arranged} so that the boxes of $\QQ_h$ are
exactly paved by those of $\QQ_{h+1}$. 

Define $z^{(h)}_\Bx$ to be the random field with propagator
$C^{(h)}_{\Bx,\Bh}$ defined as the Fourier transform of $\sum_{\V
n\in\zzz^d}\big (\fra{1}{\V p^2+\g^{-2}m^2}-\fra1{\V
p^2+m^2}\big) e^{i\V n\cdot\V p
\,L\,\g^{h}}$: so that $\f^{(\le N)}_\Bx$ and its propagator $C^{(\le
N)}_{\Bx,\Bh}$ can be represented, see \equ(1.2),\equ(1.3), as

$$\f^{(\le N)}_\Bx\=\sum_{h=1}^{N} \g^{\fra{d-2}2 h}z^{(h)}_{\g^h\Bx},
\qquad C^{(\le N)}_{\Bx,\Bh}=\sum_{h=1}^N \g^{(d-2)h}
C^{(h)}_{\g^h\Bx,\g^h\Bh}\Eq(2.2)$$
where the fields $z^{(h)}$ are independently distributed Gaussian
fields. Note that the fields $z^{(h)}$ are {\it also almost
identically distributed} because their propagator is obtained by periodizing
over the period $\g^h L$ the {\it same} function $\lis
C^{(0)}_{\Bx,\Bh}\defi\ig
\fra{e^{i\V p\cdot(\Bx-\Bh)}d\V p} {(2\p)^d} \big(\fra{1}{\V p^2
+\g^{-2}m^2}-\fra{1}{\V p^2 +m^2}\big)$:
\ie their propagator is $C^{(h)}_{\Bx,\Bh}=\sum_{\V
n\in\zzz^d} \lis C^{(0)}_{\Bx,\Bh+\g^h\V n L}$. The reason why they
are not exactly equally distributed is that the field $z^{(h)}_\Bx$ is
periodic with period $\g^h L$ rather than $L$. But proceeding with
care the sum over $\V n$ in the above expressions can be essentially
ignored: this is a little price to pay if one wants
translation invariance built in the analysis since the beginning.

The representation \equ(2.2) defines a {\it multiscale
representation} of the field $\f^{(\le N)}_\Bx$.
Smoothness properties for the field $\f^{(\le N)}_\Bx$
can be read from those of its ``components''
$z^{(h)}$. Define, for $\D\in \QQ_0$,

$$||z^{(h)}||_\D=\max_{\Bx\in\D,\Bh\in\L\atop |\Bx-\Bh|\le m^{-1}}
\big(|z^{(h)}_\Bx|+
\t\fra{|z^{(h)}_\Bx-z^{(h)}_\Bh|}{|\Bx-\Bh|^{\fra14}}\big)
\Eq(2.3)$$
and $\t$ will be chosen $\t=0$ or $\t=1$ as needed (in practice $\t=0$
if $d=2$ and $\t=1$ if $d=3$): $\t=1$ will allow to discuss some
smoothness properties of the fields which will be necessary (\eg if
$d=3$). Then the size $||z||_\D$ {\it of any field} $z^{(h)}$, for all
$h\ge1$, is estimated by

$$P(\max_{\D\subset \QQ_0} ||z||_\D\le B) \ge e^{-c\, e^{- c'\,
B^2}\,|\L|}, \qquad
P(||z||_\D\ge B_\D, \forall\D\in\DD)\le \prod_{\D\in\DD} c\, e^{-c'\,
B_\D^2}\Eq(2.4)$$
where $P$ is the Gaussian probability distribution of $z$, $\DD$ is
any collection of boxes $\D\in \QQ_0$ and $c,c'>0$ are suitable
constants. The \equ(2.4) imply in particular \equ(2.1). The estimates
\equ(2.4) follow from the Markovian nature of the Gaussian field
$z^{(h)}$, \ie from the fact that the propagator is the Green's
function of an elliptic operator (of fourth order, see the first of
\equ(1.3)), with constant coefficients which implies also the inequalities 
(fixing $\e\in(0,1)$)

$$|C^{(h)}_{\Bx,\Bh}|\=
\big|\ig z_{\Bx}z_{\Bh} \,P(d z) \big|\le\, c\, e^{-m
\,c'\,|\Bx-\Bh|},\qquad 
|C^{(h)}_{\Bx,\Bh}-C^{(h)}_{\Bx,\Bh'}|\le c \, (m|\Bh-\Bh'|)^{\e},
\Eq(2.5)$$
where $|\Bx-\Bh|$ is {\it reinterpreted} as the distance between $\Bx,\Bh$
measured over the periodic box $\g^h\L$ (hence $|\Bx-\Bh|$ differs
from the ordinary distance only if the latter is of the order of
$\g^hL$). The interpretation of \equ(2.5) is that $z^{(h)}_{\Bx}$ are
essentially {\it bounded} variables which, on scale $\sim m^{-1}$, are
essentially {\it constant} and furthermore beyond length $\sim m^{-1}$
are essentially {\it independently distributed}.

\*
\0{\it References:} [Wi72],[Ga81],[Ga85].
\*

\Section(3, Perturbation theory)
\*

The naive approach to the problem is to fix $\l_N\=\l>0$ and to
develop $Z_N(\L,f)$ or, more conveniently and equivalently,
$\fra1{|\L|}\log Z_N(\L,f)$ in powers of $\l$. If one fixes \ap
$\m_N,\n_N$ independent of $N$, however, even a formal power series is
not possible: this is trivially due to the divergence of the
coefficients of the power series, {\it already to second order} for
generic $f$ in the limit $N\to\io$. {\it Nevertheless} it is possible
to determine $\m_N(\l),\n_N(\l)$ as functions of $N$ and $\l$ so that
a formal power series exists (to all orders in $\l$): this is the key
result of {\it renormalization theory}.

To find the perturbative expansion the simplest is to use a
graphical representation of the coefficients of the power expansion in
$\l,\m_N,\n_N,f$ and the Gaussian integration rules which yield (after
a classical computation) that the coefficient of $\l^n\m_N^p
f_{\Bx_1}\ldots f_{\Bx_r}$ is obtained by considering the following
{\it graph elements}

\eqfig{230pt}{42pt}{
\ins{18pt}{10pt}{$\Bx$}
\ins{100pt}{10pt}{$ \Bx$}
\ins{160pt}{10pt}{$ \Bx$}
\ins{220pt}{10pt}{$ \Bx$}
}
{fig1}{(1)}

\0where the segments will be called {\it half lines} and the graph
elements will be called, respectively, {\it coupling} {\rm or} {\it
$\f^4$-vertex}, {\it mass vertex}, {\it vacuum vertex} and {\it
external vertex}.

The half lines of the graph elements are considered {\it distinct}
(\ie imagine a label attached to distinguish them). Then consider all
possible {\it connected} graphs $G$ obtained by first drawing,
respectively, $n,p,r$ graph elements in Fig.1, which are not vacuum
vertices, with their nodes marked by points in $\L$ named
$\Bx_1,\ldots,\Bx_n,
\Bx_{n+1},\ldots,$ $\Bx_{n+p+r}$; and form all possible graphs obtained by
attaching pairs of halph lines emerging from the vertices of the graph
elements. These are the ``nontrivial graphs''. Furthermore consider
also the single ``trvial'' graph formed just by the third graph
element and consisting of a single point. All graphs obtained in this
way are particular {\it Feynman graphs}.

Given a nontrivial graph $G$ (there are many of them) we define its {\it
value} to be the product

$$W_G(\Bx_1,\ldots,\Bx_n,
\Bx_{n+1},\ldots,\Bx_{n+p+r})=(-1)^{n+p+r}\fra{\l^n\m_N^p\prod 
f_{\Bx_{n+p+j}}}{n!p!r!}\prod_\ell C^{(\le
N)}_{\Bx_\ell,\Bh_\ell}\Eq(3.1)$$
where the last product runs over all pairs $\ell=(\Bx_\ell,\Bh_\ell)$ of half
lines of $G$ that are joined and connect two vertices labeled by points
$\Bx_\ell,\Bh_\ell$: call {\it line} of $G$ any such pair. If the graph
consists of the single vacuum vertex its value will be $\n_N$.  The
series for $\fra1{|\L|} \log Z_N(\L,f)$ is then

$$
-\n_N+\fra1{|\L|}\sum_G\ig W_G(\Bx_1,\ldots,
\Bx_{n+p+r}) 
\prod_{j=1}^{n+p+r} d\Bx_j\Eq(3.2)$$
and the integral will be called the {\it integrated graph value}. 

Suppose first that $\m_N=\n_N=0$. Then if a graph $G$ contains
subgraphs like

\eqfig{250pt}{48pt}
{\ins{-4pt}{10pt}{$\Ba$}
\ins{38pt}{10pt}{$\Bx$}
\ins{80pt}{10pt}{$\Bb$}
\ins{116pt}{10pt}{$\Ba$}
\ins{134pt}{10pt}{$\Bx$}
\ins{182pt}{10pt}{$\Bh$}
\ins{200pt}{10pt}{$\Bb$}
%
}{fig2a}{(2)}

\0the corresponding respective contribution to the integral in
\equ(3.2) (considering only the integrals over $\Bh$ and suitably
taking care of the combinatorial factors) is a factor obtained by
integrating over $\Bx$ the quantities

$$-6\,\l\, C^{(\le N)}_{\Ba\Bx}C^{(\le N)}_{\Bx\Bx}C^{(\le N)}_{\Bx\Bb}
\quad{\rm or}\quad \fra{4^2\cdot
3!}{2!}
\,\l^2\,
C^{(\le N)}_{\Ba\Bx} \ig C^{(\le
N)\,3}_{\Bx\Bh}\,C^{(\le
N)}_{\Bh\Bb}\,d\Bh
\Eq(3.3)$$
which if $d=3$ diverge as $N\to\io$ as $\g^{N}$ or, respectively, as
$N$; the second factor does not diverge in dimension $d=2$ while the
first still diverges as $N$. The divergences arise from the fact that
as $\Bx-\Bh\to\V0$ the propagator behaves as $|\Bx-\Bh|^{-N}$ if $d=3$
or as $-\log|\Bx-\Bh|$ if $d=2$, all the way until saturation occurs
at distance $|\Bx-\Bh|\simeq m^{-1}\g^{-N}$: for this reason the
latter divergences are called {\it ultraviolet divergences}.

However if we set $\m_N\ne0$ then for every graph containing a
subgraph like those in Fig.2 there is another one identical except
that the points $\Ba,\Bb$ are connected via a mass vertex, see Fig.1,
with the vertex in $\Bx$, by a line $\Ba\Bx$ and a line $\Bx\Bb$; the
new graph value receives a contribution from the mass vertex inserted
in $\Bx$ between $\Ba$ and $\Bb$ simply given by a factor
$-\m_N$. Therefore if we fix, for $d=3$,

$$\m_N=-6\,\l\,C^{(\le N)}_{\Bx\Bx}+ \fra{4^2\cdot 3!}{2}\,\l^2\,
\ig_\L C^{(\le
N)\,3}_{\Bx\Bh}\,d\Bh\defi-6\,\l\,C^{(\le N)}_{\Bx\Bx}+\d\m_N\Eq(3.5)$$
we can simply consider graphs which {\it do not contain any mass graph
element and in which there are no subgraphs like the first in Fig.2}
while the subgraphs like the second in Fig.2 do not contribute a
factor $\ig C^{(\le N)}_{\Ba\Bx}C^{(\le N)\,3}_{\Bx\Bh}C^{(\le
N)}_{\Bh\Bb}\,d\Bh$ but a {\it renormalized} factor $\ig C^{(\le
N)}_{\Ba\Bx}C^{(\le N)\,3}_{\Bx\Bh}\big(C^{(\le N)}_{\Bh\Bb}-C^{(\le
N)}_{\Bx\Bb}\big)\,d\Bh$.  If $d=2$ we only need to define $\m_N$ as
the first term in the \rhs of \equ(3.5) and we can leave the subgraphs
like the second in Fig.2 as they are (without any renormalization).

Graphs without external lines are called {\it
vacuum graphs} and there are a few such graphs which are divergent.
Namely, if $d=3$, they are the first three drawn in Fig.2';
furthermore if $\m_N$ is set to the above nonzero value a new vacuum
graph, the fourth in Fig.2', can be formed. Such graphs

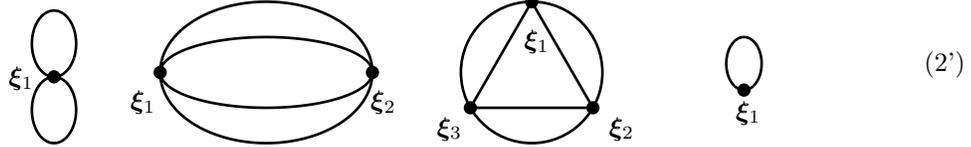
\eqfig{330pt}{55pt}
{\ins{8pt}{29pt}{$\Bx_1$}
\ins{54pt}{20pt}{$\Bx_1$}
\ins{145pt}{20pt}{$\Bx_2$}
\ins{170pt}{10pt}{$\Bx_3$}
\ins{235pt}{10pt}{$\Bx_2$}
\ins{203pt}{43pt}{$\Bx_1$}
\ins{283pt}{16pt}{$\Bx_1$}
}
{fig3d}{(2')}

\0contribute to the graph value, respectively, the addends in the sum

$$-3\,\l\, C^{(\le N)\,2}_{\Bx_1,\Bx_1}+ \fra{4!}2 \,\l^2\,\ig C^{(\le
N)\,4}_{\Bx_1\Bx_2}\,d\Bx_2  -\fra{2^3 \cdot3!^3}{3!}\l^3
\ig C^{(\le
N)\,2}_{\Bx_1\Bx_2}C^{(\le
N)\,2}_{\Bx_2\Bx_3}C^{(\le
N)\,2}_{\Bx_3\Bx_1}\,d\Bx_2 \,d\Bx_3-\m_N C^{(\le N)}_{\Bx_1\Bx_1}
\Eq(3.4)$$

\0and diverge, respectively, as $\g^{2N},\g^{N},N,\g^{2N}$ 
if $d=3$ while, if $d=2$, only the first and the last diverge, like
$N^2$.

Therefore if we fix $\n_N$ as minus the quantity in \equ(3.5) we can
disregard graphs like those in Fig.2'; if $d=2$ $\n_N$ can be defined to be
the sum of the first and last terms in \equ(3.4).

The formal series in $\l$ and $f$ thus obtained is called the {\it
renormalized series} for the field $\f^4$ in dimension $d=2$ or,
respectively, $d=3$. Note that with the given definitions and choices
of $\m_N,\n_N$ the {\it only} graphs $G$ that need to be considered to
construct the expansion in $\l$ and $f$ are formed by the first and
last graph elements in Fig.1, paying attention that the grapfs in
Fig.2' do not contribute and, if $d=3$, the graphs with subgraphs like
the second in Fig.2 have to be computed with the modification
described.

In the next section it will be shown that the above are the only
sources of divergences as $N\to\io$ and therefore the problem of
studying \equ(1.1) is solved at the level of formal power series by
the {\it subtraction} in
\equ(3.5). This also shows that giving a meaning to the
series thus obtained is likely to be much easier if $d=2$ than if
$d=3$.

The coefficients of order $k$ of the expansion in $\l$ of
$\fra1{|\L|}\log Z_N(\L,f)$ can be ordered by the number $2n$ of
vertices representing external fields: and have the form $\ig
S^{(k)}_{2n}(\Bx_1,\ldots,\Bx_{2n})\prod_{i=1}^{2n}
(f_{\Bx_i}d\Bx_i)$: the kernels $S^{(k)}_{2n}$ are the Schwinger
functions of order $2n$, see Sect.\sec(1).
\*

\0{\it Remark:} if $d=4$ the regularization at cut-off $N$ in \equ(1.2)
is not sufficient as in the subtraction procedure smoothness of the
first derivatives of the field $\f^{(\le N)}$ is necessary, while the
regularization \equ(1.2) does not even imply \equ(2.1), \ie not even
H\"older continuity. A higher regularization (\ie using a $\ch_N$ like
the square of the $\ch_N$ in \equ(1.3)).  Furthermore the subtractions
discussed in the case $d=3$ are not sufficient to generate a formal
power series and many more subtractions are needed: for instance
graphs with a subgraph like 

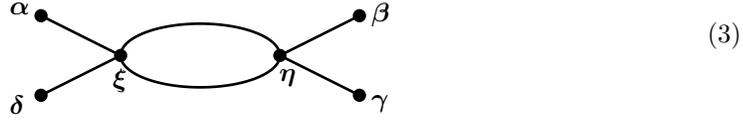
\eqfig{120pt}{40pt}
{
\ins{27pt}{10pt}{$\Bx$}
\ins{90pt}{10pt}{$\Bh$}
\ins{-12pt}{35pt}{$\Ba$}
\ins{125pt}{35pt}{$\Bb$}
\ins{125pt}{0pt}{$\Bg$}
\ins{-12pt}{0pt}{$\Bd$}
}
{fig3}{(3)}
\*

\0would give a contribution to the graph value which is a factor
$\l^2 \ell_N\defi \fra{2\cdot 6^2}{2!}\,\l^2\, \ig_\L C^{(\le
N)\,2}_{\Bx\Bh}\,d\Bh$, also divergent as $N\to\io$ proportionally to
$N$. Although this divergence could be canceled by changing $\l$ into
$\l_N=\l+\l^2 \ell_N$ the previously discussed cancellations would be
affected and a change in the value of $\m_N$ would become necessary;
furthermore the subtraction in \equ(3.5) will not be sufficient to
make finite the graphs, not even to second order in $\l$, unless a new
term $-\a_N\ig (\dpr_\Bx\f^{(\le N)}_\Bx)^2\,d\Bx$ with
$\a_N=\fra12\l^2\ig \dpr_\Bh C^{(\le N)\,3}_{\Bx\Bh}(\Bx-\Bh)^2$ is
added in the exponential in \equ(1.1). But all this will not be enough
and still new divergences, proportional to $\l^3$, will appear.

And so on indefinitely: the consequence being that it will be
necessary to define $\l_N,\m_N,\a_N,\n_N$ as formal power series in
$\l$ (with coefficients diverging as $N\to\io$) in order to obtain a
formal power series in $\l$ for \equ(1.1) in which all coefficients
have a finite limit as $N\to\io$. Thus the interpretation of the
formal renormalized series in the case $d=4$ is substantially
different and naturally harder than the cases $d=2,3$. Beyond formal
perturbation expansions the case $d=$ {\it is still an
open problem}: the most widespread conjecture is that the series cannot
be given a meaning other than setting to $0$ all coefficients of
$\l^j,\,j >0$. In other words, the conjecture claims, there should be
no nontrivial solution to the ultraviolet problem for scalar $\f^4$
fields in $d=4$. But this is far from being proved, even at a
heuristic level.  The situation is simpler if $d\ge5$: in such cases
it is impossible to find formal power series in $\l$ for
$\fra1{|\L|}\log Z_N(\L,f)$, even allowing $\l_N,\m_N,\a_N,\n_N$ to be
formal power series in $\l$ with divergent coefficients.
\*

The distinctions between the cases $d=2,3,4,>4$ explain the terminology
given to the $\f^4$-scalar field theories calling them {\it
superrinormalizable} if $d=2,3$, renormalizable if $d=4$ and {\it non
renormalizable} if $d>4$.
Since the (divergent) coefficients in the formal power series defining
$\l_N,\m_N,\a_N,\n_N$ are called {\it counterterms} the  $\f^4$-scalar
fields require finitely many counterterms (see \equ(3.5)) in the
superrenormalizable cases and infinitely many in the renormalizable
case. The nonrenormalizable cases ($d>4$) cannot be treated in a way
analogous to the renormalizable ones.
\*

\0{\it References:} [Ga85],[Fr82].
\*

\Section(4, Finiteness of the renormalized series, $d=2,3$:
{``power counting''}.)
\*

Checking that the renormalized series is well defined to all orders
is a simple {\it dimensional estimate} characteristic of many
multiscale arguments that in Physics have become familiar with the name
of ``renormalization group arguments''.

Consider a graph $G$ with $n+r$ vertices built over $n$ graph elements
with vertices $\Bx_1,\ldots,\Bx_n$ each with $4$ half lines and $r$
graph elements with vertices $\Bx_{n+1},\ldots,\Bx_{n+r}$ representing
the external fields: as remarked in Sec.\sec(3) these are the only
graphs to be considered to form the renormalized series.

Develop {\it each} propagator into a sum of propagators as in
\equ(2.2).  The graph $G$ value will, as a consequence, be represented
as a sum of values of new graphs obtained from $G$ by adding {\it
scale labels} on its lines and the value of the graph will be computed
as a product of factors in which a line joining $\Bx\Bh$ and bearing a
scale label $h$ will contribute with $C^{(h)}_{\Bx\Bh}$ replacing
$C^{(\le N)}_{\Bx\Bh}$. To avoid proliferation of symbols we shall
call the graphs obtained in this way, \ie with the scale labels
attached to each line, still $G$: no confusion should arise as
{\it we shall, henceforth, only consider graphs $G$ with each line
carrying also a scale label}.

The scale labels added on the lines of the graph $G$ allow us to
organize the vertices of $G$ into {\it clusters}: a cluster of {\it
scale $h$} consists in a maximal set of vertices (of the graph elements in the
graph) connected by lines of scale $h'\ge h$ among which one at least
has scale $h$.

It is convenient to consider the vertices of the graph elements as ``
trivial'' clusters of highest scale: conventionally call them clusters
of scale $N+1$.

The clusters can be of ``first generation'' if they contain only
trivial clusters, of ``second generation'' if they contain only
clusters which are trivial or of the first generation, and so on.

Imagine to enclose in a box the vertices of graph elements inside a
cluster of the first generation and then into a larger box the
vertices of the clusters of the second generation and so on: the set
of boxes ordered by inclusion can then be represented by a {\it rooted
tree} graph whose nodes correspond to the clusters and whose ``top
points'' are nodes representing the trivial clusters (\ie the vertices
of the graph).

If the maximum number of nodes that have to be crossed to reach a top
point of the tree starting from a node $v$ is $n_v$ ($v$ included and
the top nodes included) then the node $v$
represents a cluster of the $n_v$--th generation.  The first node before
the root is a cluster containing all vertices of $G$ and the root of
the tree will not be considered a node and it can conventionally bear the
scale label $0$: it represents symbolically the value of the graph.

For instance in Fig.4 a tree $\th$ is drawn: its nodes correspond to
clusters whose scale is indicated next to them; in the second part of
the drawing the trivial clusters as well as the clusters of the first
generation are enclosed into boxes.

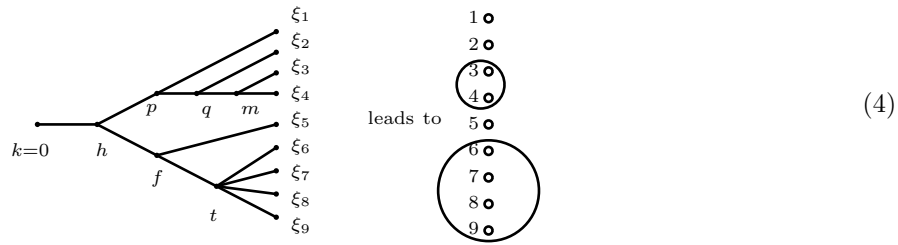
\eqfig{240pt}{80pt}
{\ins{-10pt}{28pt}{$\st k=0$}
\ins{22pt}{28pt}{$\st h$}
\ins{41pt}{43pt}{$\st p$}
\ins{62pt}{42pt}{$\st q$}
\ins{77pt}{42pt}{$\st m$}
\ins{43pt}{18pt}{$\st f$}
\ins{65pt}{3pt}{$\st t$}
\ins{96pt}{80pt}{$\st \x_1$}
\ins{96pt}{70pt}{$\st \x_2$}
\ins{96pt}{60pt}{$\st \x_3$}
\ins{96pt}{50pt}{$\st \x_4$}
\ins{96pt}{40pt}{$\st \x_5$}
\ins{96pt}{30pt}{$\st \x_6$}
\ins{96pt}{20pt}{$\st \x_7$}
\ins{96pt}{10pt}{$\st \x_8$}
\ins{96pt}{0pt} {$\st \x_9$}
\ins{125pt}{40pt}{$\st {\rm leads\ to}$}
\ins{163pt}{78pt}{$\st 1$}
\ins{163pt}{68pt}{$\st 2$}
\ins{163pt}{58pt}{$\st 3$}
\ins{163pt}{48pt}{$\st 4$}
\ins{163pt}{38pt}{$\st 5$}
\ins{163pt}{28pt}{$\st 6$}
\ins{163pt}{18pt}{$\st 7$}
\ins{163pt}{8pt}{$\st 8$}
\ins{163pt}{-2pt}{$\st 9$}
}
{fig4}{(4)}
\*

Then consider the next generation clusters, \ie the clusters which
only contain clusters of the first generation or trivial ones, and
draw boxes enclosing all the graph vertices that can be reached from
each of them by descending the tree, \etc. Fig.5 represents all boxes
(of any generation) correspondinf to the nodes of the tree in Fig.4

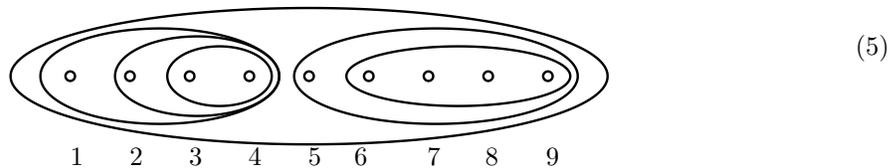
\eqfig{210pt}{70pt}
{
\ins{0pt} {0pt}{$1$}
\ins{22.5pt}{0pt}{$2$}
\ins{45pt}{0pt}{$3$}
\ins{67.5pt}{0pt}{$4$}
\ins{90pt}{0pt}{$5$}
\ins{107.5pt}{0pt}{$6$}
\ins{135pt}{0pt}{$7$}
\ins{157pt}{0pt}{$8$}
\ins{180pt}{0pt}{$9$}
}
{fig5}{(5)}

\*

\0The representations of the clusters of a graph $G$ by a tree or by
hierarchically ordered boxes (see Fig.4 and Fig.5) are completely
equivalent provided inside each box not representing a top point of
the tree the scale $h_v$ of the corresponding cluster $v$ is marked.
For instance in the case of Fig.5 one gets

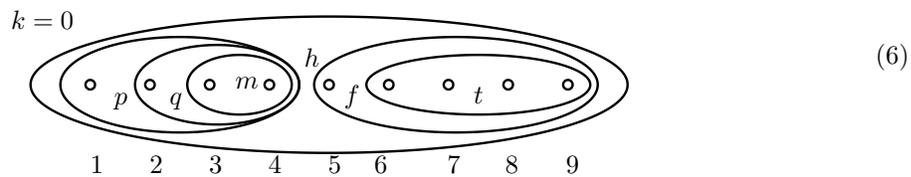
\eqfig{210pt}{70pt}
{
\ins{-30pt}{55pt}{$k=0$}
\ins{55pt}{30pt}{$m$}
\ins{30pt}{24pt}{$q$}
\ins{9pt}{24pt}{$p$}
\ins{81pt}{41pt}{$h$}
\ins{96pt}{27pt}{$f$}
\ins{145pt}{26pt}{$t$}
\ins{0pt} {0pt}{$1$}
\ins{22.5pt}{0pt}{$2$}
\ins{45pt}{0pt}{$3$}
\ins{67.5pt}{0pt}{$4$}
\ins{90pt}{0pt}{$5$}
\ins{107.5pt}{0pt}{$6$}
\ins{135pt}{0pt}{$7$}
\ins{157pt}{0pt}{$8$}
\ins{180pt}{0pt}{$9$}
}
{fig6}{(6)}

\*
\0By construction if two top points $\Bx$ and $\Bh$ are inside the same
box $b_v$ of scale $h_v$ but not in inner boxes then there is a path
of graph lines joining $\Bx$ and $\Bh$ all of which have scales $\ge
h_v$ and one at least has scale $h_v$.

Given a graph $G$ fix one of its points $\Bx_1$ (say) and integrate
the absolute value of the graph over the positions of the remaining
points.  The exponential decay of the propagators implies that if a
point $\Bh$ is linked to a point $\Bh'$ by a line of scale $h$ the
integration over the position of $\Bh'$ is essentially constrained to
extend only over a distance $\g^{-h}m^{-1}$. Furthermore the maximum
size of the propagator associated with a line of scale $h$ is bounded
proportionally to $\g^{(d-2)h}$. Therefore, recalling that $|f_{\Bx}|$
is supposed bounded by $1$, the mentioned integral can be immediately
bounded by

$$ \fra{\l^n}{n!r!} \, C^{n+r}\,I\,\defi \,\fra{\l^n \, C^{n+r}}{n!r!}
\,\prod_\ell \g^{\fra{d-2}2\,h_\ell} \prod_v \g^{-d
\,h_v\,(s_v-1)}\Eq(4.1)$$
where, $C$ being a suitable constant, the first product is over the
half lines $\ell$ composing the graph lines and the second is over the
tree nodes (\ie over the clusters of the graph $G$), $s_v$ is the
number of subclusters contained in the cluster $v$ but not in inner
clusters; and in \equ(4.1) the scale of a half line $\ell$ is $h_\ell$
if $\ell$ is paired with another half line to form a line $\ell$ (in the
graph $G$) of scale label $h_\ell$.

Denoting by $v'$ the cluster immediately containing $v$ in $G$, by
$n^{inner}_{v}$ the number of half lines in the cluster $v$, by
$n_v,r_v$ the numbers of graph elements of the first type or of the
fourth type in Fig.1 with vertices in the cluster $v$, and denoting by
$n^e_v$ the number of lines which are not in the cluster $v$ but
have one extreme on a vertex in $v$ (``lines external to $v$''), the
identities ($k=0$)

$$\eqalign{
&\sum_{v>root} (h_v-k)(s_v-1)\=\sum_{v>root}(h_v-h_{v'})(n_v+r_v-1),\cr
&\sum_{v>root} (h_v-k)\,n^{inner}_{v}\=
\sum_{v>root}(h_v-h_{v'})\,\widetilde n^{inner}_{v},\qquad\hbox{with}\cr
&\widetilde n^{inner}_{v}\defi \,4n_{v}+r_v-n^e_{v},\cr
}\Eq(4.2)$$
hold, so that the estimate \equ(4.1) can be elaborated into

$$I\le \prod_{v>r}\g^{-\r_v\,(h_v-h_{v'})},
\qquad \r_v\defi-d+(4-d)n_v+r_v \fra{d+2}2+\fra{d-2}2 n^e_{v}\Eq(4.3)$$
where $h_{v'}=k=0$ if $v$ is the first nontrivial node (\ie $v'=root$),
and an estimate of the integral of the absolute value of the graphs
$G$ with given tree structure but different scale labels is
proportional to $\sum_{\{h_v\}} I<\io$ if (and only if) $\r_v>0,
\forall v$.

{\it But} there may be clusters $v$ with only two external lines
$n^e_v=2$ and two graph vertices inside: for which $\r_v=0$. However
this can happen {\it only if $d=3$} and in {\it only one case}: namely
if the graph $G$ contains a subgraph of the second type in Fig.2 and
the three intermediate lines form a cluster $v$ of scale $h_v$ while
the other two lines are external to it: hence on scale $h'>h$. In this
case one has to remember that the subtraction in Sec.\sec(3) has led
to a modification of the contribution of such a subgraph to the value
of the graph (integrated over the position labels of the vertices). As
discussed in Sec.\sec(3) the change amounts at replacing the
propagator $C^{(h')}_{\Bh,\Bb}$ by
$C^{(h')}_{\Bh,\Bb}-C^{(h')}_{\Bx,\Bb}$.

This improves, in \equ(4.3), the estimate of the contribution of the
line joining $\Bh$ to $\Bb$ from being proportional to $\ig C^{(\le
h_v)\,3}_{\Bx\Bh} C^{(\le h')}_{\Bh\Bb}d\Bh$ to being proportional to
$\ig C^{(\le h_v)\,3}_{\Bx\Bh} (C^{(\le h')}_{\Bh\Bb}-C^{(\le
h')}_{\Bx\Bb})d\Bh$; and this changes the contribution of the line
$\Bh\Bb$ from $\g^{(d-2)h'}$ to $\ig e^{-m\g^{h_v}|\Bx-\Bh|}
(\g^{h'}|\Bx-\Bh|)^\fra12d\Bh$ because $C^{(h')}$ is regular on scale
$\g^{-h'}m^{-1}$, see \equ(2.5) with $\e=\fra12$.

Since $\Bx,\Bh$ are in a cluster of higher scale $h_v$ this means that
the estimate is improved by $\g^{-\fra12(h_v-h')}$. In terms of the
final estimate this means that $\r_v$ in \equ(4.3) can be improved to
$\lis\r_v=\r_v+\fra12$ for the clusters for which $\r_v=0$. Hence the
integrated value of the graph $G$ (after taking also into account the
integration over the initially selected vertex $\Bx_1$, trivially
giving a further factor $|\L|$ by translation invariance), and summed
over the possible scale labels is bounded
proportionally to $|\L|\sum_{\{h_v\}} I<\io$ once the estimate of
$I$ is improved as described.

Note that the graphs contributing to the perturbation series for
$\fra1{|\L|}\log Z_N(\L,f)$ to order $\l^n$ are finitely many because
the number $r$ of external vertices is $r\le 2n+2$ (since graphs must
be connected). Hence the perturbation series is finite to all orders
in $\l$.

The above is the renormalizability proof of the scalar $\f^4$-fields
in dimension $d=2,3$. The theory is renormalizable even if $d=4$ as
mentioned in the remark at the end of Sec.\sec(3). The analysis would be
very similar to the above: it is just a little more involved power
counting argument.
\*
\0{\it References:} [He66], Sect. 8 and 16 in [Ga85].
\*

\Section(5, Asymptotic freedom {($d=2,3$)}. Heuristic analysis.)
\*

Finiteness to all orders of the perturbation expansions is by no
means sufficient to prove the existence of the ultraviolet limit for
$Z_N(\L,f)$ or for $\fra1{|\L|}\log{Z_N(\L,f)}$: and
\ap it might not even be necessary.  For this purpose the first step
is to check uniform (upper and lower) boundedness of $Z_N(\L,f)$ as
$N\to\io$.

The reason behind the validity of a bound $e^{|\L| E_-(\l,f)}\le
{Z_N(\L,f)}\le e^{|\L| E_+(\l,f)}$ with $E_\pm(\l,f)$ cut-off
independent has been made very clear after the introduction of the
renormalization group methods in field theory. The approach studies
the integral $Z_N(\L,f)$, recursively, decomposing the field $\f^{(\le
N)}_\Bx$ into its regular components $z^{(h)}_\Bx$, see \equ(2.2), and
integrating first over $z^{(N)}$, then over $z^{(N-1)}$ and so on.

The idea emerges naturally if the potential $V_N$ in
\equ(1.1), \equ(1.4) is written in terms of the ``normalized'' variables
$X^{(N)}_\Bx\defi\g^{-\fra{d-2}2 N}\f^{(\le N)}_\Bx$, see \equ(2.1);
here if $d=2$ the factor $\g^{\fra{d-2}2 N}$ is interpreted as
$N^{\fra12}$.

The key remark is that {\it as far as the integration over the small scale
component $z^{(N)}$ is concerned} the field $X^{(N)}_\Bx$ is a sum of two
fields of size of order $1$ (statistically), $X^{(N)}_\Bx\= z^{(N)}_{\g^N
\Bx}+ \g^{-\fra{d-2}2}X^{(N-1)}_\Bx$ (if $d=2$ this becomes
$X^{(N)}_\Bx\= \fra1{N^{\fra12}}z^{(N)}_{\g^N\Bx}+
\fra{(N-1)^{\fra12}}{N^{\fra12}} X^{(N-1)}_\Bx$)
and it can be considered to be smooth on scale $m^{-1}\g^{-N}$ (also
statistically). Hence {\it approximately constant} and of size of
order $O(1)$ on the small cubes $\D$ of volume $\g^{-d N}m^{-d}$ of
the pavement $\QQ_N$ introduced before \equ(2.2); at the same time it
can be considered to take (statistically) {\it independent values} on
different cubes of $\QQ_N$. This is suggested by the inequalities
\equ(2.3),\equ(2.4),\equ(2.5).

Therefore it is natural to decompose the potential $V_N$, see
\equ(1.5), as a sum over the small cubes $\D$ of volume $\g^{-d
N}m^{-d}$ of the pavement $\QQ_N$ as (see \equ(3.5) for the definition
of $\m_N,\n_N$), taking henceforth
$m=1$,

$$V_N(z^{(N)})\defi-\sum_{\D\in\QQ_N} \g^{-Nd}\ig_\D\Big(
\l \g^{2(d-2)N} X^{(N)\,4}_\Bx+\m_N \g^{(d-2)N}
X^{(N)\,2}_\Bx+\n_N
+f_\Bx \g^{\fra{d-2}2N}X^{(N)}_\Bx \Big)\fra{d\Bx}{|\D|}\Eq(5.1)$$
where $\g^{(d-2)N}$ is interpreted as $N$ if $d=2$. Hence if $d=3$ it is

$$V_N(z^{(N)})\defi-\sum_{\D\in\QQ_N} \g^{-N}\ig_\D\Big(
\l  X^{(N)\,4}_\Bx+\lis\m_N X^{(N)\,2}_\Bx+\lis\n_N
+f_\Bx \g^{-\fra32N}X^{(N)}_\Bx\Big) \fra{d\Bx}{|\D|}\Eq(5.2)$$
where $\lis\m_N\defi (-6\l c_N+\l^2N \g^{-N}c'_N)$, $\lis\n_N\defi 3\l
c_N^2+\l^2\g^{-N}b_N+\l^3N \g^{-2N}b'_N$, and $c_N,c'_N,b_N,b'_N$,
computable from \equ(3.4),\equ(3.5), admit a limit as $N\to\io$. While
if $d=2$ it is

$$V_N(z^{(N)})\defi-\sum_{\D\in\QQ_N} N^2\g^{-2N}\ig_\D \Big(
\l  X^{(N)\,4}_\Bx+\lis\m_N X^{(N)\,2}_\Bx+\lis\n_N
+f_\Bx N^{-\fra32}X^{(N)}_\Bx \Big)
\fra{d\Bx}{|\D|}\Eq(5.3)$$
where $\lis\m_N\defi-6\l c_N$ and $\lis\n_N=3\l c_N^2$ and $c_N$,
computable from \equ(3.3), admits a limit as $N\to\io$.
\*

The fields $z^{(N)}$ and $X^{(N-1)}$ can be considered constant over
boxes $\D\in\QQ_N$: $z^{(N)}_\Bx=s_\D, X^{(N-1)}_\Bx= x_\D$ for
$\Bx\in\D$ and the $s_\D$ can be considered statistically independent
on the scale of the lattice $\QQ_N$.  

{\it Therefore \equ(5.2),\equ(5.3) show that integration over
$z^{(N)}$ in the integral defining $Z_N(\L,f)$ is not too different
from the computation of a partition function of a lattice continuous
spin model in which the ``spins'' are $s_\D$ and, most important,
interact extremely weakly if $N$ is large.} In fact the coupling
constants are of order of a power of $|X^{(N-1)}|$ times $O(\g^{-N})$
if $d=3$ ($O(N^2\g^{-2N})$ if $d=2$), or of order
$O(\g^{-\fra{d+2}2N}\max|f_\Bx|)$, {\it no matter} how large $\l$ and
$f$.

This says that the smallest scale fields are {\it extremely weakly
coupled}. The fields $X^{(N-1)}$ can be regarded as external fields of
size that will be called $B_{N-1}$, of order $1$ or even allowed to
grow with a power of $N$, see \equ(2.1). Their presence in $V_N$
does not affect the size of the couplings, as far as the analysis of
the integral over $z^{(N)}$ is concerned, because the couplings remain
{\it exponentially small} in $N$, see \equ(5.2),\equ(5.3), being at
worst multiplied by a {\it power} of $B_{N-1}$, \ie changed by a
factor which is a power of $N$.
\*

The smallness of the coupling at small scale is a property
called {\it asymptotic freedom}. Once fields and coordinates
are ``correctly scaled'' the real size of the coupling becomes
manifest, \ie it is extremely small and the addends in $V_N$
proportional to the ``counterterms'' $\m_N,\n_N$, which looked
divergent when the fields were not properly scaled, are in fact of the
same order or much smaller than the main $\f^4$-term.

Therefore the integration over $z^{(N)}$ can be, heuristically,
performed by techniques well established in statistical mechanics (\ie
by straightforward perturbation expansions): {\it at least if the
field $X^{(\le N-1)}_\Bx$ is smooth and bounded}, as prescribed by
\equ(2.1), with $B=B_{N-1}$ growing as a power of $N$. 
In this case, denoting symbolically the integration over $z^{(N)}$ by
$P$ or by $\media{\ldots}$, {\it it can be expected} that it  should give

$$\ig e^{V_N} dP(z^{(N)})\=e^{V_{j;N-1}+\lis \RR(j,N) |\L|}
\Eq(5.4)$$
where $V_{j;N-1}$ is the Taylor expansion of $\log\ig e^{V_N} dP(z^{(N)})$
in powers of $\l$ (hence essentially in the very small parameter $\l
\g^{-(4-d)N}$) truncated at order $j$, \ie

$$\eqalignno{
V_{1;N-1}=&[\media{V_N}]^{\le 1},\qquad
V_{2;N-1}=\Big[\media{V_N}+
\fra{(\media{V_N^2}-\media{V_N}^2)}{2!}\Big]^{\le 2},&\eq(5.5)\cr
V_{3;N-1}=&\Big[
\media{V_N}+\fra{(\media{V_N^2}-\media{V_N}^2)}{2!}+
\fra {\big(\media{V_N(\media{V_N^2}-\media{V_N}^2)}
-\media{V_N}(\media{V_N^2}-\media{V_N}^2)\big)}{3!}\Big]^{\le 3},\ 
\ldots\cr}$$
where $[\cdot]^{\le j}$ denotes truncation to order $j$ in
$\l$, and $\lis \RR(j,N)$ is a remainder (depending on $\f^{(\le
N-1)}_\Bx$) which can be {\it  expected} to be estimated by

$$
|\lis \RR(j,N)|\le \RR(j,N)\defi C_j  B_N^{4j}(\l\,N^2\,
\g^{-(4-d)N})^{j+1}\g^{dN},\qquad{\rm}\  {\rm for}\
d=2,3\Eq(5.6)$$
for suitable constants $C_j$, \ie a remainder estimated by the
$(j+1)$-th power of the coupling times the number of boxes of scale
$N$ in $\L$. The relations
\equ(5.4),\equ(5.5),\equ(5.6) result from a naive Taylor expansion (in $\l$ of
the $\log \ig e^{V_N} dP(z^{(N)})$, taking into account that, in $V_N$
as a function of $z^{(N)}$, the $z^{(N)}$'s appear multiplied by
quantities at most of size $\le \l\g^{4-d}N^2 B_N^3$, by
\equ(5.2),\equ(5.3) if $|X^{(N-1)}|\le B_{N-1}$). In a statistical
mechanics model for a lattice spin system such a calculation of $Z_N$
would lead to a mean field equation of state once the remainder was
neglected.

The peculiarity of field theory is that a relation like
\equ(5.4),\equ(5.6) has to be applied again to $V_{j;N-1}$ to perform
the integration over $z^{(N-1)}$ and define $V_{j;N-2}$ and, then,
again to $V_{j;N-2}$ ...  Therefore it will be essential to perform
the integral in \equ(5.4) to an order (in $\l$) high enough so that
the bound $\RR(j,N)$ can be summed over $N$: this requires (see
\equ(5.6)) an explict calculation of \equ(5.5) pushed at least to order $j=1$ if
$d=2$ or to order $j=3$ if $d=3$ and a check that the resulting
$V_{j;N-1}$ can still be interpreted as low coupling spin model so
that \equ(5.4) can be iterated with $N-1$ replacing $N$ and then with
$N-2$ replacing $N-1$,....

The first necessary check towards a proof of the discussed heuristic
``expectations'' is that, defining recursively $V_{j;h}$ from
$V_{j,h+1}$ for $h=N-1,\ldots,1,0$ by \equ(5.5) with $V_N$ replaced by
$V_{j;h+1}$ and $V_{j;N-1}$ replaced by $V_{j;h}$, the couplings
between the variables $z^{(h)}$ do not become 'worse' than those
discussed in the case $h=N$. Furthermore the field $\f^{(\le
N-1)}_\Bx$ has a high probability of satisfying \equ(2.1), {\it but
fluctuations are possible}: hence the $\RR$-estimate has to be
combined with another one dealing with the large fluctuations of
$X^{(N-1)}_\Bx$ which has to be shown to be ``not worse''..
\*

\0{\it References:} [Ga78],[Ga85],[BG95].
\*

\Section(6, Effective potentials  and their scale {(in)}dependence.)
\*

To analyze the first problem mentioned at the end of Sec.\sec(5), {\it
define} $V_{j;h}$ by \equ(5.5) with $V_N$ replaced by $V_{j;h+1}$ for
$h=N-1,N-2,\ldots,0$.  The quantities $V_{j;h}$, which are called {\it
effective potentials} on scale $h$ (and order $j$), turn out to be in
a natural sense {\it scale independent}: this is a consequence of
renormalizability, realized by Wilson as a much more general property
which can be checked, in the very special cases considered here with
$d=2,3$, at fixed $j$ by induction, and in the superrinormalizable
models considered here it requires only an elementary computation of a
few Gaussian integrals as the case $j=3$ (or {\it even} $j=1$ if
$d=2$) is already sufficient for our purposes.

It can, also, be (more easily) proved for general $j$ by a dimensional
argument parallel to the one presented in Sec.\sec(4) to check
finiteness of the renormalized series. The derivation is elementary
but {\it it should be stressed that, again, it is possible only
because of the special choice of the counterterms $\m_N,\n_N$}. If
$d=3$ the boundedness and smoothness of the fields $\f^{(\le h)}$ and
$z^{(h)}$ expressed by the second of \equ(2.1) and of \equ(2.5) is
essential; while if $d=2$ the smoothness is not necessary.

The structure of $V_{j;h}$ is conveniently expressed in terms of the
fields $X^{(h)}_\Bx$, as a sum of three terms $V_{h}^{(rel)}$
(standing for ``relevant'' part), $V_{h}^{(irr)}$ (standing for
``irrelevant'' part) and a ``field independent'' part $E(j,h)|\L|$. 

The relevant part in $d=2$ is simply of the form \equ(5.3) with $h$
replacing $N$: call it $V^{(rel,1)}_h$.  If $d=3$ it is given by
\equ(5.2) with $h$ replacing $N$ plus, for $h<N$, a second 
``nonlocal'' term $V^{(rel,2)}_h\defi
\fra{4^2\, 3!}{2!\,2!}\,\l^2 \ig (C_{\Bh\Bh'}^{(\le
h)\,3}-C_{\Bh\Bh'}^{(\le N)\,3}) (\f^{(\le h)}_{\Bh}-\f^{(\le
h)}_{\Bh'})^2d\Bh d\Bh'$ which is conveniently expressed in terms of a
``non local'' field $Y^{(h)}_{\Bh\Bh'}\defi \fra{\f^{(\le
h)}_{\Bh}-\f^{(\le h)}_{\Bh'}}{(\g^h|\Bh-\Bh'|)^{\fra14}}$ as
$V_h^{(rel)}=V_h^{(rel,1)}+V_h^{(rel,2)}$ with

$$
V^{(rel,2)}_h\defi -\l^2 \g^{-2h}\sum_{\D,\D'\in\QQ_h}\ig_{\D\times\D'}
Y^{(h)\,2}_{\Bh\Bh'}\,A^{(h)}_{\Bh\Bh'}\, e^{-c'\g^h|\Bh-\Bh'|}\,
 \fra{d\Bh d\Bh'}{|\D|\,|\D'|}
\Eq(6.1)$$
where $0<a\le\Big(\fra{A^{(h)}_{\Bh\Bh'}}{(\g^h|\Bh-\Bh'|)^{3-\fra12}}\Big)_N <
a'$, with $a,a',c'>0$ and the subscript $N$ means that the expression
in parenthesis ``saturates at scale $N$'', \ie it becomes
$\g^{(3-\fra12)(h-N)}$ as $|\Bh-\Bh'|\to0$.

The expression \equ(6.1) is not the full part of the potential
$V_{j;h}$ which is of second order in the fields: there are several
other contributions which are collected below as ``irrelevant''.

It should be stressed that {\it irrelevant} is a traditional technical
term: by no means it should suggest ``neglegibility''. On the contrary it
could be maintained that the whole purpose of the theory is to study
the irrelevant terms. A better word to designate the irrelevant part
of the potential would be {\it driven part} as its behavior is
``controlled'' by the relevant part. The Schwinger functions are
simply related to the irrelevant terms.

The irrelevant part of the effective potential can be expressed as a
finite sum of integrals of monomials in the fields $X^{(h)}_\Bx$ if
$d=2$, or in the fields $X^{(h)}_\Bx${\it and} $Y^{(h)}_{\Bh\Bh'}$ if
$d=3$, which can be written as $V^{(irr)}_{j;h}$ given by

$$\ig (\prod_{k=1}^p X^{(h)\,n_k}_{\Bx_k}
\prod_{k'=1}^q Y^{(h)\,n'_{k'}}_{\Bh_{k'}\Bh'_{k'}})
\,e^{-\g^{h} c'
d(\Bx_1,\ldots,\Bh'_q)} \l^{n}\g^{-h t}
W(\Bx_1\ldots,\Bh'_q)
\prod_{k=1}^p\fra{d\Bx_k}{|\D_k|} \prod_{k'=1}^q
\fra{d\Bh_{k'}d\Bh'_{k'}}{|\D^1_{k'}|\,|\D^2_{k'}|}
\Eq(6.2)$$
with the integral extended to products
$\D_1\times\ldots\D_p\times\ldots\times (\D^1_{q}\times\D^2_{q})$ of
boxes $\D\in\QQ_h$, and $d(\Bx_1,\ldots,\Bh'_q)$ is the length of the
shortest tree graph that connects all the $p+2q>0$ points, the
exponents $n,t$ are $\ge2$ and $t$ is $\ge3$ if $q>0$; the kernel $W$
depends on all coordinates $\Bx_1\ldots,\Bh'_q$ and it is bounded
above by $C_j \prod_{k'=1}^q A_{\Bh_{k'}\Bh'_{k'}}$ for some $C_j$; the
sums $\sum n_{k}+\sum n'_{k'}$ cannot exceed $4j$. The test
functions $f$ do not appear in \equ(6.2) because by assumption they
are bounded by $1$: but $W$ depends on the $f$'s as well.

The field independent part is simply the value of $\log Z_N(\L,f)$
computed by the perturbation analysis in Sec.\sec(3) {\it up to order
$j$ in $\l$ but using as propagator $(C^{(\le N)}-C^{(\le h)})$}: thus
$E(j,h)$ is a constant depending on $N$ but uniformly bounded as
$N\to\io$ (because of the renormalizability proved in Sec.\sec(3)).

If $d=2$ there is no need to introduce the nonlocal fields $Y^{(h)}$
and in \equ(6.2) one can simply take $q=0$, and the relevant part also
can be expressed by omitting the term $V^{(rel,2)}_h$ in \equ(6.1):
unlike the $d=3$ case the estimate on the kernels $W$ by an
$N$-independent $C_j$ holds uniformly in $h$ without having to
introduce $Y$. For $d=2$ it will therefore
be supposed that $V_h^{(rel,2)}\=0$ in \equ(6.1) and $q=0$ in \equ(6.2). 

{\it It is not necessary to have more informations on the structure of
$V_{j;h}$} even though one can find simple graphical rules, closely
related to the ones in Sec.\sec(3), to construct the coefficients $W$
in full detail. The $W$ depend, of course, on $h$ but the uniformity
of the bound on $W$ is the only relevant property and in this sense
the effective potentials are said to be (almost) ``scale independent''.

The above bounds on the irrelevant part can be checked by an
elementary direct computation if $j\le3$: in spite of its ``elemetary
character'' the uniformity in $h\le N$ is a result ultimately playing
an essential role in the theory together with the dominance of the
relevant part over the irrelevant one which, once the fields are
properly scaled, is ``much smaller'' (by a factor of order $\g^{-h}$,
see \equ(6.2)).
\*

\0{\it Remarks:} 
(1) Checking scale independence for $j=1$ is just checking that
$\ig P(d z^{(h)}) V_{1;h}=V_{1;h-1}$. Note that $V_{1;h}\defi\ig_\L
\l(\f^{(\le h)\,4}_\Bx-6 C^{(\le h)}_{\V0\V0}\f^{(\le
h)\,2}_\Bx+3C^{(\le h)\,2}_{\V0\V0})d\Bx$; hence calling $:\f^{(\le
h)\,4}_\Bx:$ the polynomial in the integral ({\it Wick's monomial of
order $4$}) this is an elementary Gaussian integral (``martingale
property of Wick monomials''). Note the essential role of the
counterterms. For $j>1$ the computation is similar but it involves
higher order polynomials (up to $4j$) and the distinction between
$d=2$ and $d=3$ becomes important.

\0(2) $V_{j;0}$ contains
only the field independent part $E(j,0)|\L|$ which is just a number
(as there are no fields of scale $0$): by the above definitions it is
{\it identical} to the perturbative expansion truncated to $j$--th
order in $\l$ of $\log Z_N(\L,f)$, well defined as discussed in
Sec.\sec(3),\sec(4).
\*

\Section(7, Nonperturbative renormalization: small fields)
\*

Having introduced the notion of effective potential $V_{j;h}$, of
order $j$ and scale $h$, satisfying the bounds (described after
\equ(6.2)) on the kernels $W$ representing it, the problem is to
estimate the remainder in \equ(5.4) and find its relation with the
value \equ(5.6) given by the heuristic Taylor expansion. Assume $\l<1$
to avoid distinguishing this case from that with $\l\ge1$ which would
lead to very similar estimates but to different $\l$-dependence on
some constants.

Define $\ch_B(z^{(h)})=1$ if $||z^{(h)}||_\D\le B h^2$ for all
$\D\in\QQ_h$, see \equ(2.3), and $0$ otherwise; then the following
lemma holds:
\*

\0{\bf Lemma 1:} {\it Let  $||X^{(h)}||_{\D}$
be defined as \equ(2.3) with $z$ replaced by $X$ and suppose
$||X^{(h)}||_{\D}\le B h^{4}$ for all $\D$ then, for all $j\ge1$, it
is

$$\ig e^{V_{j;h+1}} \ch_B(z^{(h+1)})\,dP(z^{(h+1)})
\,=\,e^{\,V_{j;h}\,+\,\RR'(j,{h+1})\,|\L|}\Eq(7.1)$$
with, for suitable constants $c_-,c'_-$, $|\RR'_-(j;h+1)|\le
\RR_-(j;h+1)\defi 
\RR(j;h+1)+ c_- e^{-c_-'
B^2 (h+1)^2}$ and $\RR(j;h+1)$ given by \equ(5.6) with $h+1$ in place
of $N$.}
\*

Since $Z_N(\L,f)\ge \ig e^{V_N}\prod_{h=1}^N \ch_B(z^{(h)})\, P(d
z^{(h)}) $ {\it this immediately gives a lower bound on} $E=\fra1{|\L|}\log
Z_N(\L,f)$: in fact if $\ch_B(||z^{(h')}||)=1$ for $h'=1,\ldots,h$
then $||X^{(h)}||_\D\le c\,B h^{'4}$ for some $c$ so that, by
recursive application of lemma 1, $Z_N(\L,f)\ge
e^{V_{j,0}-\sum_{h=1}^N
\RR_-(j,{h})|\L|}$. 
By the remark at the end of Sec.\sec(6), given $j$ the lower bound on
$E$ just described agrees with the perturbation expansion of
$E=\fra1{|\L|} \log Z_N(\L,f)$ truncated to order $j$ (in $\l$) up to
an error $\sum_{h=1}^N\RR_-(j,{h})$.
\*

\0{\it Remark:} The problem solved by lemma 1 is called
the {\it small fields problem}. The proof of the lemma is a simple
Taylor expansion in $\l\g^{-h}$ if $d=3$ or in $\l h^2\g^{-2h}$ if
$d=2$ to order $j$ (in $\l$). The constraint on $z^{(h+1)}$ makes the
integrations over $z^{(h+1)}$, necessary to compute $V_{j;h}$ from
$V_{j;h+1}$, not Gaussian. But the {\it tail estimates} \equ(2.4),
together with  the Markov property of the ditributionof $z^{(h)}$ can
be used to estimate the difference with respect to the Gaussian
unconstrained integrations of $z^{(h+1)}$: and the result
is the addition of the small ``tail error'' changing $\RR$ into
$\RR_-$. The estimate of the main part of the remainder $\RR$ would be
obvious if the fields $z^{(h)}$ were independent on boxes of scale
$\g^{-h}$: they are not independent but they are Markovian and
the estimate can be done by taking into account the Markov property.
\*
\0{\it References:} [Wi70],[Wi72],[Ga78],[Ga81],[BCGNPOS78],[Ga85].
\*

\Section(8, Nonperturbative
renormalization: large fields, ultraviolet stability)
\*

The small fields estimates are {\it not sufficient} to obtain
ultraviolet stability: to control the cases in which $|X^{(h)}_\Bx|> B
h^{4}$ for some $\Bx$ or some $h$, or $|Y^{(h)}_{\Bx\Bh}|> B
h^{4}$ for some $|\Bx-\Bh|<\g^{-h}$, a further idea is necessary and it
rests on making use of the assumption that $\l>0$ which, in a sense to
be determined, should suppress the contribution to the integral
defining $Z_N(\L,f)$ coming from very large values of the
field. Assume also $\l<1$ for the same reasons advanced in
Sec.\sec(6).

Consider first $d=2$.  Let $\DD_N$ be the ``large field region'' where
$|X^{(N)}_\Bx|> B N^{4}$ and let $V_N(\L/\DD_N)$ be the integral defining
the potential in
\equ(5.3) extended to the region $\L/\DD_N$, complement of
$\DD_N$. This region is typically {\it very} irregular (and random as
$X$ itself is random with distribution $P_N$).

An upper bound on the integral defining $Z_N(\L,f)$ is obtained by
simply replacing $e^{V_N}$ by $e^{V_N(\L/\DD_N)}$ because in $\DD_N$
the first term in the integrand in \equ(5.3) is $\le -\l N^2\g^{2N} (B
N^4)<0$ and it overwhelmingly dominates on the remaining terms whose
value is bounded by a similar expression with a smaller power of $N$.
Then if $\EE^c\defi \L/\EE$ denotes the complement in $\L$ of a set
$\EE\subset \L$:
\*

{\bf Lemma 2:} {\it Let $d=2$. Define $V_h(\DD^c_h)$,to be given
by the expression
\equ(5.4)  with the integrals extending over $\D_j/\DD_h$ 
and define $\RR(j,h+1)$ by \equ(5.6).  Then

$$\ig e^{V_{h+1}(\DD^c_{h+1})}\,dP(z^{(h+1)})
=e^{V_h(\DD^c_h)+\lis\RR_+(j,{h+1})|\L|}\Eq(8.1)$$
where $|\lis\RR_+(j,{h+1}|\le \RR_+(j,{h+1}\defi \RR(j;h+1)+ c_+ e^{-c'_+
B^2 (h+1)^2}$ with suitable $c_+,c'_+$.}
\*

\0{\it Remark:} 
Lemma 2 is genuinely not perturbative and making essential use of the
positivity of $\l$. Below the analysis of the proof of the lemma,
which consists essentially in its reduction to Lemma 1, is described
in detail.  It is perhaps the most interesting part and the core of
the theory of the proof that truncating the expansion in $\l$ of
$\fra1{|\L|}\log Z_N(\L,f)$ to order $j$ gives as a result an estimate
{\it exact} to order $\l^{j+1}$ of $\fra1{|\L|}\log Z_N(\L,f)$.
\*

Let $R_N$ be the cubes $\D\in\QQ_{N}$ in which there is at least one
point $\Bx$ where $|z^{(N)}_\Bx|\ge B N^2$. 
By definition, the region $\DD_N/\DD_{N-1}$ is
covered by $R_N$.

Remark that in the region $\DD_{N-1}/R_N$ the field
$X^{(N-1)}$ is large but $z_N$ is not large so that $X^{(N)}$ is still
very large: this is so because the bounds set to define the regions
$\DD$ and $R$ are quite different being $B N^4$ and $B N^2$
respectively. Hence if a point is in $\DD_{N-1}$ and not in $R_N$ then
the field $X^{(N)}$ must be of the order $\gg B N^3$. Therefore {\it by
positivity} of the $\l\f^{(\le N)\,4}_\Bx$ term (which dominates all
other terms so that $ V^{(N)}(\f^{(\le N)}_\Bx)<0$ for
$\Bx\in\DD_N\cup(\DD_{N-1}/R_N)$) we can replace $V_N(\DD^c_N)$ by
$V((\DD_N\cup(\DD_{N-1}/R_N))^c)$, for the purpose of obtaining an upper
bound.

Furthermore modulo a suitable correction it is possible to replace
$V((\DD_N\cup(\DD_{N-1}/R_N))^c)$ by $V((\DD_{N-1}\cup R_N)^c)$: because the
integrand in $V_N$ is bounded below by $-b \l \g^{-2N}N^2$ if $d=2$ (by
$-b\l \g^{-N} $ if $d=3$), for some $b$, so that the points in $R_N$ can
at most lower $V((\DD_N\cup(\DD_{N-1}/R_N))^c)$ by $-b\l N^2
\g^{-(4-d)N}\,\#(R_N)$ if $\# R_N$ is the number of boxes of $\QQ_N$
in $R_N$ and $V(\f_\Bx)$ is bounded {\it below by its minimum}: thus
$V((\DD_{N-1}\cup R_N)^c)+b\l N^2
\g^{(4-d)N}\,\#(R_N)$ is an upper bound to $V((\DD_N\cup(\DD_{N-1}/R_N))^c)$.

In the complement of $\DD_{N-1}\cup R_N$ all fields are ``small''; if
$X^{(N-1)}$ and $R_N$ are fixed this region is not random (as a
function of $z^{(N)}$) any more. Therefore if $X^{(N-1)},R_N$ are
fixed the integration over $z^{(N)}$, {\it conditioned to having
$z^{(N)}$ fixed (and large) in the region $R_N$}, is performed by
means of the same argument necessary to prove lemma 1 (essentially a
Taylor expansion in $\l\g^{-(4-d)N}$). The large size of $z^{(N)}$ in
$R_N$ does not affect too much the result because on the boundary of
$R_N$ the field $z^{(N)}$ is $\le B N^2$ (recalling that $z^{(N)}$ is
continuous) and since the variable $z^{(N)}$ is Markovian the boundary
effect decays exponentially from the boundary $\dpr R_N$: it adds a
quantity that can be shown to be bounded by the number of boxes in
$R_N$ on the boundary of $R_N$, hence by $\# R_N$, times $b' (N-1)^2
\g^{-(4-d)} (B(N-1)^4)^4 $ for some $b'$.

The result of the integration over $z^{(N)}$ of
$e^{V_N((\DD_{N}\cup(\DD_{N-1}/R_N))^c)}$ {\it conditioned to the large
field values of $z^{(N)}$ in $R_N$} leads to an upper bound on $\ig
e^{V_N} P(dz^{(N)})$ as

$$
\sum_{R_N} e^{V_{j;N-1}(\DD^c_{N-1})+\RR(j,{N})|\L|}
\prod_{\D\in R_N} \Big(c \,e^{-c' (B N^2)^2 } e^{+c''\l \g^{-(4-d)N}
N^2 (B N^4)^4}\Big)^{\# R_N}\Eq(8.2)$$
where $c,c',c''$ are
suitable constants: this is explained as follows.
\*

\0(i) Taylor expansion (in $\l$) of the integral
$e^{V_N((\DD_{N-1}\cup R_N)^c)+b\l N^2 \g^{-(4-d)N}\#(R_N)}$ (which, by
construction, is an upper bound on $e^{V_N(\DD^c_N)}$) with respect to
the field $z^{(N)}$, conditioned to be fixed and large in $R_N$, would
lead to an upper bound as $e^{V_{j;N-1}((\DD_{N-1}\cup
R_N)^c)+\RR'(j,{N})|\L|+b''\l(B N^4)^4 \g^{(4-d)N}\,\#(R_N)}$ with $\RR'$
equal to \equ(5.6) possibly with some $C'_j$ replacing $C_j$.  The
second exponential in the \rhs of \equ(8.2) arises partly from the
above correction $b''\l(B N^4)^4 \g^{-(4-d)N}\,\#(R_N)$ and partly from
a contribution of similar form explained in (iii) below.

\0(ii) Integration over the large conditioning fields fixed in $R_N$
is controlled by the second estimate in \equ(2.4) (the {\it tail
estimate}): the first factors in parenthesis is
the tail estimate just mentioned, \ie the probability that $z^{(N)}$
is large in the region $R_N$. The second factor is only partly explained in
(i) above.

\0(iii) Without further estimates the bound \equ(8.2)
would contain $V_{j;N-1}((\DD_{N-1}\cup R_N)^c)$ rather than
$V_{j;N-1}(\DD^c_{N-1})$. Hence there is the need to 
change the potential $V_{j;N-1}((\DD_{N-1}\cup R_N)^c)$ by ``reintroducing''
the contribution due to the fields in $R_N/\DD_{N-1}$ in order to
reconstruct $V_{j;N-1}(\DD^c_{N-1})$. Reintroducing this part of the
potential costs a quantity like $b'\l N^2 \g^{(4-d)N} (B N^4)^4\#(R_N)$
(because the reintroduction occurs in the region $R_N/\DD_{N-1}$ which
is covered by $R_N$ and in such points the field $X^{(N-1)}_\Bx$ is
not large, being bounded by $B (N-1)^4$); so that their contribution to
the effective potential is still dominated by the $\f^4$ term and
therefore by $ \g^{-(4-d)N}$ times a power of $BN^4$ times the volume
of $R_N$ (in units $\g^{-N}$, \ie $\#R_N$). All this is taken care of
by suitably fixing $c''$.
\*

Note that the sum over $R_N$ of \equ(8.2) is $(1+ c \,e^{-c' B^2 N^4 }
e^{+c''\l \g^{-(4-d)N} N^2(B N^4)^4})^{\g^{d N}|\L|}$ (because $\L$
contains $|\L|\g^{dN}$ cubes of $\QQ_N$) hence it is bounded above by
$e^{c_+ e^{-c'_+ B^2 N^2}}$ for suitably defined $c_+,c'_+$.

The {\it same argument} can be repeated for $V_{j;h}(\DD^c_h)$ with any
$h$ if $V_{j;h}(\DD^c_h)$ is defined by the sum over $\D$'s in $\QQ_h$
of the same integrals as those in
\equ(6.1),\equ(6.2) with $\D_j/\DD_h$ replacing $\D_j$ in the integration
domains.

Applying lemma 1 and lemma 2 recursively (with $j\ge3$) it follows
that there exist {\it $N$-independent} upper and lower bounds $E_\pm \,|\L|$
on $\log Z(\L,f)$ of the form $V_{j;0}\pm\sum_{h=1}^\io
(\RR(j,h)+c_\pm e^{-c'_\pm B^2 h^2})|\L|$ for $c_\pm,c'_\pm>0$ suitably
chosen and $\l$--independent for $\l<1$.  By the remark at the end of
Sec.\sec(6), given $j$ the bounds just described agree with the
perturbation expansion $E(j,0)|\L|\=V_{j;0}$ of $\log Z(\L,f)$
truncated to order $j$ (in $\l$) up to the remainders
$\pm\sum_{h=1}^N\RR_\pm(j,{h})$. Hence if $B$ is chosen proportional
to $\log_+ \l^{-1}\defi \log(e+\l^{-1})$ the upper and lower bounds coincide
to order $j$ in $\l$ with the value obtained by truncating to order
$j$ the perturbative series.  

The latter remark is important as it implies not only that the bounds
are finite (by Sec.\sec(3)) but also that $\fra1{|\L|}\log Z(\L,f)$ is
{\it not quadratic in $f$}: already to order $1$ in $\l$ it is quartic
in $f$ (containing a term equal to $-\l (\ig
C_{\Bx,\V0}\,f_\Bx\,d\Bx)^4$). Thus the outline of the proof of lemma 2,
which together with lemma 1 forms the core of the analysis of the
ultraviolet stability for $d=2$, is completed.
\*

If $d=3$ more care is needed because (very mild) smoothness, like the
considered H\"older continuity with exponent $\fra14$, of $z,X$ is
necessary to obtain the key scale independence property discussed in
Sec.\sec(6): therefore the natural measure of the size of $z^{(h)}$
and $X^{(h)}$ in a box $\D\in\QQ_h$ {\it is no longer the maximum of
$|z^{(h)}_\Bx|$ or of $|X^{(h)}_\Bx|$}.  The region $\DD_h$ becomes
more involved as it has to consist of the points $\Bx$ where
$|X^{(h)}_\Bx|>B h^4$ and of the pairs $\Bh,\Bh'$ where
$|Y_{\Bh,\Bh'}|\=\fra{|X^{(h)}_\Bh-X^{(h)}_{\Bh'}|}
{(\g^h|\Bh-\Bh'|)^{\fra14}}>Bh^4$: \ie it is not just a subset of
$\L$.

However, if $d=3$, the relevant part {\it also contains the negative
term $V^{(rel,2)}$, see \equ(6.1)}: and since it dominates over all
other terms which contain a $Y$-field (because their coupling are
smaller by about $\g^{-h}$) the argument given for $d=2$ can be
adapted to the new situation.  Two regions $\DD_h^1,\DD_h^2$ will be
defined: the first consists of all the points $\Bx$ where
$|X^{(h)}_\Bx|>B h^4$ and the second of all the pairs $\Bh,\Bh'$ where
$|Y^{(h)}_{\Bh,\Bh'}|>Bh^4$.  The region $R_h$ will be the collection
of all $\D\in\QQ_h$ where $||z^{(h)}||_\D>B h^2$.  Then $V(\DD_h^c)$
will be defined as the sum of the integrals in \equ(6.1), \equ(6.2)
with the integrals over $\Bx_i$ further restricted to $\Bx_i\not\in
\DD^1_h$ and those over the pairs $\Bh_i,\Bh'_i$ are further
restricted to $(\Bh_i,\Bh'_i)\not\in\DD^2_h $. With the new settings
lemma 2 can be proved also for $d=3$ along the same lines as in the
$d=2$ case.
\*

\0{\it References:} [Wi70],[Wi72],[BCGNPOS78],[Ga81].

\*

\Section(9, Ultraviolet limit, infrared behavior and other applications)
\*

The results on the ultraviolet stability are nonperturbative, as no
assumption is made on the size of $\l$ (the assumption $\l<1$ has been
imposed in Sec.\sec(7),\sec(8) only to obtain simpler expressions for
the $\l$--dependence of various constants): nevertheless the
multiscale analyis has allowed us to use perturbative techniques (\ie
the Taylor expansion in lemmata 1,2) to find the solution. The latter
procedure is the essence of the renormalization group methods: they aim
at reducing a difficult multiscale problem to a sequence of simple
single scale problems. Of course in most cases it is difficult to
implement the approach and the scalar quantum fields in dimension
$2,3$ are among the simplest examples. The analysis of the {\it beta
function} and of the {\it running couplings}, which appear in
essentially all renormalization group applications, does not play a
role here (or, better, their role is so inessential that it has even
been possible to avoid mentioning them). This makes the models
somewhat special from the renormalization group viewpoint: the running
couplings at length scale $h$, if introduced, would tend exponentially
to $0$ as $h\to\io$; unlike what happens in the most interesting
renormalization group applications in which they either tend to zero
only as powers of $h$ or do not tend to zero at all.

The multiscale analysis method, \ie the renormalization group method,
in a form close to the one discussed here has been applied very often
since its introduction in Physics and it has led to the solution of
several important problems. The following is a not exhaustive list
together with a few open questions.

(1) The arguments just discussed imply with minor extra work that
   $Z_N(\L,f)$ as $N\to\io$ not only admit uniform upper and lower
   bounds but also that the limit as $N\to\io$ actually exists and it
   is a $C^\io$ function of $\l,f$. Its $\l$ and $f$--derivatives at
   $\l=0$ and $f=0$ are given by the formal perturbation
   calculation. In some cases it is even possible to show that the
   formal series for $Z_N(\L,f)$ in powers of $\l$ is Borel
   summable. An interesting question is to explore the possibility of
   an ultraviolet stability proof which is exclusively based on the
   perturbation expansion without having recourse to the probabilistic
   methods in the analysis.

(2) The problem of removing the infrared cut--off (\ie $\L\to\io$) is
    in a sense more a problem of statistical mechanics. In fact it can
    be solved for $d=2,3$ by a typical technique used in statistical
    mechanics, the {\it cluster expansion}. This is not intended to
    mean that it is technically an easy task: understanding its
    connection with the low density expansions and the possibility of
    using such techniques has been a major achievement that is not
    discussed here.

(3) The third problem mentioned in the introduction: \ie checking the
    axioms so that the theory could be interpreted as a quantum field
    theory is a difficult problem which required important efforts to
    control and which is not analyzed here. An introduction to it can
    be its analysis in the $d=2$ case.

(4) Also the problem of keeping the ultraviolet cut--off and removing
    the infrared cut--off while the parameter $m^2$ in the propagator
    approaches $0$ is a very interesting problem related to many
    questions in statistical mechanics at the critical point.

(5) Field theory methods can be applied to various statistical
    mechanics problems away from criticality: particularly interesting
    is the theory of the neutral Coulomb gas and of the dipole gas in
    two dimensions.

(6) The methods can be applied to Fermi systems in field theory as
    well as in equilibrium statistical mechanics.  The understanding
    of the ground state in not exactly soluble models of spinless
    fermions in $1$ dimension at small coupling is one of the
    results.  And via the trasfer matrix theory it has led to the
    understanding of nontrivial critical behavior in $2$-dimensional
    models that are not exactly soluble (like Ising next nearest
    neighbor or Ashkin--Teller model). Fermi systems are of particular
    interest also because in their analysis the large fields problem
    is absent, but this great technical advantage is somewhat offset
    by the anticommutation properties of the Fermionic fields: which do
    not allow us to employ probabilistic techniques in the estimates.

(7) An outstanding open problem is whether the scalar $\f^4$-theory is
    possible and nontrivial in dimension $d=4$: this is a case of a
    renormalizable not asymptotically free theory. The conjecture that
    many support is that the theory is necessarily trivial (\ie the
    function $Z_N(\L,f)$ becomes necessarily a Gaussian in the limit
    $N\to\io$).

(8) Very interesting problems can be found in the study of highly
    symmetric quantum fields: gauge invariance presents serious
    difficulties to be studied (rigorously or even heuristically)
    because in its naive forms it is incompatible with
    regularizations. Rigorous treatments have been in some cases
    possible and in few cases it has been shown that the naive
    treatment is not only not rigorous but it leads to incorrect
    results.

(9) In connection with item (8) an outstanding problem is to understand
    relativistic pure gauge Higgs-fields in dimension $d=4$: the latter have
    been shown to be ultraviolet stable but the result has not been
    followed by the study of the infrared  limit.

(10) The classical gauge theory problem is {\it quantum
    electrodynamics}, QED, in dimension $4$: it is a renormalizable
    theory (taking into account gauge invariance) and its perturbative
    series truncated after the first few orders give results that can
    be directly confronted with experience, giving very accurate
    predictions. Nevertheless the model is widely believed to be
    incomplete: in the sense that, if treated rigorously, the result
    would be a field describing free non interacting assemblies of
    photons and electrons. It is believed that QED can make sense only
    if embedded in a model with more fields, representing other
    particles (\eg the {\it standard model}), which would influence
    the behavior of the electromagnetic field by providing an
    effective ultraviolet cut-off high enough for not alterig the
    predictions on the observations on the time and energy scales on
    which present (and, possibly, future over a long time span)
    experiments are performed. In dimension $d=3$ QED is
    superrenormalizable, once the gauge symmetry is properly taken
    into account, and it can be studied with the techniques described
    above for the scalar fields in the corresponding dimension.
\*

In general constructive quantum field theory seems to be deep in a
crisis: the few solutions that have been found concern very special
problems and are very demanding technically; the results obtained have
often not been considered to contribute appreciably to any
``progress''. And many consider that the work dedicated to the subject
is not worth the results that one can even hope to obtain. Therefore
in recent years attempts have been made to follow other paths: an
attitude that in the past usually did not lead to great achievements
but that is always tempting and worth pursuing because the rare major
progresses made in Physics resulted precisely by such changes of
attitude, leaving aside developments requiring work which was too
technical and possibly hopeless: just to mention an important case one
can recall quantum mechanics which disposed of all attempts at
understanding the observed atomic levels quantization on the basis of
refined developments of classical electromagnetism.
\*

\0{\it References:} [Gu72],[GJS73],[Si74],
[BCGNPOS78],[GJ81],[Ai82],[Fr82],[GK83],
[GK85],[GK85b],\\[Ba83],
[BGM04],[GM05].
\*

\0{\titolo References}

\*\0[Ai82] Aizenman, M., 
{\it Geometric analysis of {$\f^4$}--fields and Ising models},
  Communications in mathematical Physics, {\bf86}, 1--48, 1982.

\*\0[Ba83] Balaban, T.: {\it {$(Higgs)_{3,2}$} quantum fields in a 
  finite volume: III. Renormalization}, Communications in mathematical
  Physics, {\bf88}, 411--445, 1983.

\*\0[BCGNPOS78] Benfatto, G., Cassandro, M., Gallavotti, G., Nicol\`o, F.,
  Presutti, E., Olivieri, E., and Scacciatelli, E. {\it Some
  probabilistic techniques in field theory}, Communications in
  Mathematical Phys\-ics {\bf 59}, 143--166, 1978. And {\it
  Ultraviolet stability in euclidean scalar field theories},
  Communications in Mathematical Physics {\bf 71}, 95--130, 1980.

\*\0[BGM04] Benfatto, G., Giuliani, A., Mastropietro, V.:
  {\it Low temperature analysis of two dimensional Fermi systems with
  symmetric Fermi surface}, Annales Henry Poincar\'e, {\bf4}, 137-193,
  2003.

\*\0[BG95] Benfatto, G. Gallavotti, G.:  {\it Renormalization group}, 
  p. 1--143, Princeton U. Press, 1995.

\*\0[DR81] De Calan, C., and Rivasseau, V. {\it Local existence of the Borel
  transform in euclidean $\phi^4_4$}, Communications in Mathematical Physics
  {\bf 82}, 69--100, 1981.

\*\0[Fr82] Fr\"ohlich, J.: {\it On   the triviality of $\l\phi^4_d$ 
  theories and the approach to the critical point in 
  $d\lower1pt\hbox{\kern5pt\raise4pt\hbox{$\st %
  >$}\kern-10pt\lower1pt\hbox{$\st (=)$}}4$
  dimensions}, Nuclear Physics {\bf B200}, 281--296, 1982.

\*\0[Ga78] Gallavotti, G.: {\it Some aspects of renormalization problems in
  statistical mechanics}, Memorie dell' Accademia dei Lincei {\bf 15},
  23--59, 1978.

\*\0[Ga81] Gallavotti, G.: {\it Elliptic operators and Gaussian processes},
  In ``Aspects statistiques et aspects physiques des processus
  Gaussiens", Colloques Internat. C.N.R.S, St. Flour", p. 349--360, 1981.

\*\0[Ga85] Gallavotti, G.: {\it
  Renormalization theory and ultraviolet stability via renormalization
  group methods}, Reviews of Modern Physics {\bf 57}, 471--569, 1985.

\*\0[GK83] Gawedzky, K., and Kupiainen, A.: {\it Block spin
  renormalization group for dipole gas and $(\dpr\phi)^4$}, Annals of
  Physics {\bf 147}, 198--243, 1983.

\*\0[GK85] Gawedzky, K., and Kupiainen, A.: {\it Gross-Neveu model
  through convergent perturbation expansion}, Communications in
  Mathematical Physics, {\bf 102}, 1--30, 1985.

\*\0[GK85b] Gawedzky, K., and Kupiainen, A.:
  {\it Massless lattice $\phi^4_4$ theory: Rigorous control of a
  renormalizable asymptotically free model}, Communications in Mathematical
  Physics {\bf 99}, 197--252, 1985.

\*\0[GM05] Giuliani, A., Mastropietro, V.: {\it Anomalous Universality
in the Anisotropic Ashkin-Teller Model}, Communications in Mathematical
  Physics, {\bf 256}, 681 - 735, 2005.

\*\0[GJS73] Glimm, J., Jaffe,  A., Spencer,T.:
  in {\sl Constructive Field theory}, ed. G. Velo, A. Wightman,
  Lecture Notes in Physics, Springer--Verlag, {\bf 25}, 132--242,
  1973.

\*\0[GJ81] Glimm, J., and Jaffe, A.: {\it Quantum Physics}, Springer--Verlag,
  1981.

\*\0[Gu72] Guerra, F.: {\it Uniqueness of the vacuum energy density and Van Hove
  phenomena in the infinite volume limit for two-dimensional
  self-coupled Bose fields}, Physical Review Letters {\bf 28},
  1213--1215, 1972.

\*\0[He66] Hepp, K.: {\it Th\'eorie de la r\'enormalization},  
  Lecture Notes in Physics, {\bf2}, Springer, 1966.

\*\0[Ne66] Nelson, E.: {\it A quartic interaction in two dimensions}, in 
  {\sl Mathematical Theory of elementary particles}, ed. R Goodman,
  I. Segal, 69--??, M.I.T, Cambridge, 1966.

\*\0[OS73] Osterwalder, K., Schrader, R.:  {\it Axioms for Euclidean
   Green's functions}, Communications in mathematical physics, {\bf
   31}, 83--112, 1973.

\*\0[Si74] Simon, B.: {\it The $P(\f)_2$ Euclidean (quantum) field
  theory}, Princeton University Press, 1974.

\*\0[SW64] Streater, R.F., Wightman, A.S.: {\it 
  PCT, spin, statistics and all that}, Benjamin-Cummings, 1964,
  reprinted Princeton U. Press, 2000.

\*\0[WG65] Wightman, A.S., G\"arding, L.: {\it Fields as
  operator-valued distributions in relativistic quantum theory}, Arkiv
  f\"or Fysik {\bf 28}, 129--189, 1965.

\*\0[Wi70] Wilson, K.G.: {\it Model of coupling constant renormalization},
Physical Review D, {\bf2}, 1438--???, 1970.

\*\0[Wi72] Wilson, K. G. {\it Renormalization of a scalar
  field in strong coupling}, Physical Review, {\bf D6}, 419--426, 1972.
\*

\0email: {\tt giovanni.gallavotti@roma1.infn.it}\\
\0web: {\tt http://ipparco.roma1.infn.it}\\
\0mail: INFN, Fisica, Roma1, P.le Moro 2, 00185 Roma.

\end